\documentstyle[12pt,psfig]{article}
\def\build#1_#2^#3{\mathrel{\mathop{\kern 0pt#1}\limits_{#2}^{#3}}}

\def\pmb#1{\setbox0=\hbox{#1}
\kern-.025em\copy0\kern-\wd0
\kern.05em\copy0\kern-\wd0
\kern-.025em\raise.0433em\box0}

\begin{document}

\title{The massive $CP^{N-1}$ model for frustrated spin systems}
\author{
P. Azaria \thanks{Laboratoire de Physique Th\'eorique des
    Liquides, Universit\'e P. et M. Curie,
4 Place Jussieu, 75252
    Paris Cedex. URA 765 of CNRS},
P. Lecheminant $ ^* $ \thanks{Groupe
    de Physique Statistique, Universit\'e de Cergy-Pontoise,
    95806 Cergy-Pontoise Cedex, France.} and
D. Mouhanna\thanks{Laboratoire de Physique Th\'eorique et Hautes Energies,
Laboratoire associ\'e au CNRS URA280, Universit\'e P. et
M. Curie, 4 Place Jussieu, 75252 Paris Cedex 05, France.}}
\maketitle

\begin{abstract}

We study the classical ${SU(N)\otimes U(1)/SU(N-1)\otimes U(1)}$ Non Linear
Sigma
model which is the continuous low energy effective field theory for $N$
component  frustrated spin systems. The $\beta$ functions for the  two coupling
constants of this model
are calculated around two dimensions at two loop order in a low temperature
expansion. Our study is completed by a large $N$ analysis of the model. The
$\beta$ functions for the coupling constants and the mass gap are calculated in
all dimensions between 2 and 4 at order ${1/N}$. As a main result we show that
the standard procedure at the basis of the $1/N$ expansion leads to results
that
partially contradict those of the weak coupling analysis.  We finally present
the procedure that reconciles the weak coupling and large $N$ analysis, giving
a
consistent picture of the expected scaling of frustrated magnets.

\end{abstract}

preprint PAR-LPTHE 95-34

\newpage

\section{INTRODUCTION}

The Renormalization Group (RG) approach to Critical Phenomena has provided one
of the greatest success of Theoretical Physics in the past twenty
years$^{\cite{wilson}}$.
The most important concept that emerged is that of effective
field theory: the recognition that, whatever complicated the microscopic
physics of a system is, its low energy, long distance properties can be well
described by a
much simpler and universal theory. The concept of low energy
effective theory is particularly transparent in Statistical Mechanics, through
the celebrated idea of block spin  procedure which allows to generate a
long distance effective action by integrating out the short distance degrees of
freedom of a microscopic spin system defined at the scale of some lattice size
$a=\Lambda^{-1}$. In practice, one has to have recourse to approximate methods
among which
the quantum field theoretical approach has been one of the most prolific and
successful one$^{\cite{brezin6,zinn}}$. In the latter, the block spin procedure
is bypassed by a
continuous limit in which is made  the choice of the fluctuating fields
together with operators satisfying the criterion
of perturbative renormalizability. The resulting renormalizable action is then
seen as the  low energy effective theory, i.e. at scale  $p\ll \Lambda$, of
the microscopic theory since non renormalizable - or irrelevant - terms
correspond
to those that are discarded by taking the long distance limit $\Lambda
\rightarrow \infty$$^{\cite{polchinski2}}$.  Critical and, more generally, long
distance quantities are then obtained from perturbative expansions of this low
energy theory. As
well known, the consistency of the perturbative approach relies on the
existence of a fixed point, around which the strength of the interactions and,
in particular, that of non renormalizable interactions can be carefully
controlled. But the main task is then to ensure that the results obtained so
far persist beyond perturbation theory. This generally consists in comparing
results obtained from different perturbative approaches. A celebrated
example for which such an approach has been particularly successful is the
Heisenberg ferromagnet$^{\cite{wilson,brezin6,polyakov,brezin,brezin1}}$.

However, many issues remain open in the context of more involved systems among
which are frustrated spin systems (for a review see ${\cite{diep2}}$). Indeed,
the effect of competing interactions on the classical Heisenberg
antiferromagnet  has been the subject of a great interest, especially focused
on the question of the universality class for its phase transition in three
dimensions. Previous studies based on the Non Linear Sigma (NL$\sigma$) model
approach by Azaria et al. in the classical$^{\cite{aza1,aza4}}$ as well as in
the quantum$^{\cite{aza3,aza5}}$ case  have revealed the existence of a non
trivial UV fixed point with $O(4)$ symmetry which governs the phase transition
above two
dimensions. But no stable fixed point has been found in the $\epsilon=4-d$
expansion of the Landau-Ginzburg-Wilson effective action so that a first order
phase transition is expected around four dimensions$^{\cite{garel,bailin}}$.
This
situation is even
more confused when one takes into account of experimental results that do not
show any evidence of universality$^{\cite{hirakawa1,takeda,itoh}}$ and those
obtained with Monte Carlo
simulations that are in favour of a continuous phase transition in $d=3$ with
exponents that differ from those of the $O(4)$
universality class$^{\cite{kawamura1,diep}}$.
Various scenarios have been proposed to account for these results but none of
them is fully satisfactory.
In any case, at the center of the debate is the
existence of the $O(4)$ fixed point found in the NL$\sigma$ model analysis.
More
precisely, the very question  is  to what extent can we believe on the results
obtained with help of  the NL$\sigma$ model beyond perturbation theory.

The validity of the low temperature expansion at the basis of the
NL$\sigma$ model approach has been intensively questioned. The possibility
that massive or irrelevant operators as well as non perturbative configurations
 drastically affect the RG properties of frustrated and non frustrated systems
is not excluded$^{\cite{wegner2,castilla,zumbach3}}$. In this work we would
like to deal with the specific situation of frustrated spin  systems. At the
heart of the problem  is the existence,  in their  RG  phase diagram, of
antagonistic fixed points  that result from the competition between different
kinds of order.
The main consequence of the presence of several fixed points is that the status
of
some operator may change: being relevant with respect to a given fixed point,
it  may become
irrelevant with respect to another one. In the low temperature expansion made
around two dimensions, such an operator remains marginal  so that one can
describe
this cross-over by means of a renormalizable theory interpolating
between the different fixed points of the theory. However, when changing
the perturbative scheme the power counting  changes and there is no guarantee
that irrelevant operators do not destabilize the $O(4)$ fixed point or, at
least,
reduce its basin of attraction so that the $O(4)$ scaling behavior
becomes hardly observable. With this respect, one may question
the existence of a universal theory that interpolates between the different
fixed points of the phase diagram. In order
to investigate the reliability of the perturbative approach in such a
multi-fixed point situation we shall, in the following,
study extensively the NL$\sigma$ model relevant for frustrated
systems by means of low temperature and large $N$ expansions.

In the second section of this paper, we present the
effective action that describes the low energy properties of frustrated spin
systems which is the  ${SU(N)\otimes U(1)/SU(N-1)\otimes U(1)}$ NL$\sigma$
model.
This model differs from that used by Azaria et al.
in their investigation  of the low temperature properties of frustrated
spin systems by the fact that the fundamental fields, in terms of which the
theory is written, span a representation of $SU(N)$ rather than of $O(N)$.
The reason we shall focus on this model will be explained in this section.

In the third section, we proceed to an extensive analysis of the
RG properties of the ${SU(N)\otimes U(1)/SU(N-1)\otimes U(1)}$
NL$\sigma$ model in a  double expansion in
powers of the temperature $T$ and $\epsilon=d-2$.
We give the $\beta$ functions for the two coupling constants entering
in this model, calculated  at two loop order. The RG
flow is solved and analyzed in $d=2$ and $d=2+\epsilon$ dimensions.
We show that the $\beta$ functions and the correlation length  $\xi$
interpolate between two fixed points, one with the $CP^{N-1}$ symmetry,
associated with high energy
degrees of freedom, the other one with the $O(2N)$ symmetry, associated with
the
long distance behavior,  which drives the finite temperature phase transition
of frustrated spin systems in $d>2$. As a main result, we show that
a new  scale $\Xi$, distinct from the correlation length, is generated
in the RG flow. This scale, that characterizes the cross-over between low and
high energy scaling behaviors, appears as the main feature of the low
temperature RG analysis of frustrated models.

It is the purpose of the fourth section of this work to address the question of
the validity  of the weak coupling analysis beyond the
formal expansion in powers of $T$ and $\epsilon$. This is why we study the
${SU(N)\otimes U(1)/SU(N-1)\otimes U(1)}$ NL$\sigma$ model in an expansion in
powers of ${1/N}$. There appears here an important point, related to the way
the
cross-over length scale $\Xi$ takes place in this approach. Indeed, in the weak
coupling analysis, this length scale is identified in the RG flow analysis but
is absent of the perturbative calculation which is completely free of any scale
dependence. In
contrast, in the ${1/N}$ expansion the situation is different and this length
scale defines itself the way to perform the perturbative expansion. As a
result we show that,
according to the value of $\Xi$, one has to perform different expansions using
different effective field theories in the vinicity of the two - $CP^{N-1}$ and
$O(2N)$ - fixed points. Then, the $\beta$ functions for the coupling constants
and the mass gap are calculated in all dimensions between 2 and 4 at order
${1/N}$. A careful comparison between the low temperature
and large $N$ expansions shows that there is an agreement of the RG
quantities in the very neighborhood of the fixed points. In the last section of
this paper we finally discuss the implications of our findings on the physics
of frustrated spin systems.

\section{THE MASSIVE ${\bf CP^{N-1}}$ MODEL}

The most striking feature with frustrated systems is that long wavelength
fluctuations are different, in nature, from the short distance ones. This is
typically the case with the Heisenberg antiferromagnetic model on the
triangular lattice. Starting with Heisenberg spins on the lattice, the strong
correlations due to geometrical frustration induce a local order in which the
spins adopt a $120^{o}$ structure on each elementary cell. Whereas the
fluctuating field at the scale of the lattice size $a$ is a vector, it turns
out to be, because of frustration, a rotation matrix $R$ of $SO(3)$ at larger
scales. It has been shown by Dombre and Read$^{\cite{dombre1}}$ that,
as the consequence of the
matrix character of the order parameter, the relevant action for frustrated
Heisenberg models is a NL$\sigma$ model defined on the homogeneous space $O(3)
\otimes O(2)/O(2)$. This model has been studied in the past with help of a low
temperature expansion. As a main result of this study is the existence of a
stable fixed point  in the neighborhood of two dimensions with increased
symmetry $O(3) \otimes O(3)/O(3) \sim O(4)/O(3)$$^{\cite{aza1,aza4}}$. Our aim
being to investigate
the reliability of the low temperature expansion, we need a generalization of
this model that allows a $1/N$ expansion. One of them is obtained when the
action derived by Dombre and Read is written
in terms of $N$ component  real basic
fields. This leads to the $O(N) \otimes O(2)/O(N-2) \otimes O(2)$ NL$\sigma$
model which has been studied by Azaria et al.$^{\cite{aza1,aza4}}$.
Another option is to parametrize the long distance fluctuations with help
of $N$ component complex fields of $SU(N)$. Doing so we obtain the
$SU(N)\otimes U(1)/SU(N-1)\otimes U(1)$ NL$\sigma$ model which we shall
extensively study in this paper. The reason to favour this model is
threefold:  first, although we shall be interested in the classical case in the
following, it appears that the $SU(N)\otimes U(1)/SU(N-1)\otimes U(1)$
NL$\sigma$ model is more appropriate to investigate the quantum case. Indeed,
there are reasons to believe that the low energy excitations in the disordered
phase of quantum frustrated spin systems are spin one half deconfined
excitations $z$ called spinons$^{\cite{sachdev}}$.
On the other hand, this model has a  more natural
large $N$ expansion than the $O(N) \otimes O(2)/O(N-2) \otimes O(2)$ version.
Indeed, as it will be shown below, it exhibits the phenomenon of enlarge
symmetry at the fixed point
for any value of $N$ whereas this property is lost for the $O(N)\otimes
O(2)/O(N-2) \otimes O(2)$ NL$\sigma$ model which has
no special symmetry properties for $ N > 3 $.
Note finally that the $SU(2)\otimes U(1)/ U(1)$  and  $O(3) \otimes O(2)/O(2)$
NL$\sigma$ models, which are relevant for the case of Heisenberg spins, differ
by their topological properties. Indeed, $SO(3)$ having a non trivial first
homotopy
group, there exist point and line defects in  two and three dimensions
respectively
which are not taken into account in the $SU(2)$ formulation.
It is not excluded that these topological defects drastically affect the
physics of frustrated spin systems$^{\cite{kawamura3}}$. We shall make however
the hypothesis that it is not the case and concentrate ourselves on the
multi-fixed point situation problem, leaving for future work the understanding
of the
role of topology.

The partition function of the $SU(N)\otimes U(1)/SU(N-1)\otimes U(1)$
NL$\sigma$ model writes$^{\cite{samuel,campos6,chubukov1,chubukov2}}$:
\begin{equation}
Z= \int D{z^{\dagger}} Dz\ \delta{\left(z^{\dagger} z-1\right)}
 \exp \left[ - \cal{S} \right]
\label{fonctpart}
\end{equation}
with:
\begin{equation}
{\cal S} = \int d^{d} x \left[ \frac{\Lambda^{d-2}}{2f_{1}}
\partial_{\mu} z^{\dagger} \partial_{\mu} z
+ \frac{\Lambda^{d-2}}{2f_{2}}
\left( z^{\dagger} \stackrel{\leftrightarrow}{\partial}_{\mu} z \right)^2
\right]\ .
\label{action1}
\end{equation}
Here $z$ is a $N$ component complex vector submitted to the
constraint $z^{\dagger} z =1$,  $\Lambda$ being the inverse of the lattice
spacing $a$ of the underlying spin system. The action ($\ref{action1}$) is
invariant under the action of left $SU(N)$ and right $U(1)$ global
transformations:
\begin{equation}
z \rightarrow U \ z \ e^{{\displaystyle{ i \theta}}},
\ \ \ U \in SU(N) \ \  \hbox{and} \ \ \theta \in I\!\!R\ .
\end{equation}
The ground state of (\ref{action1}) is:
%$z_0 = (0,....,0,1)$
\begin{equation}
z_0 = \left( \begin{array}{c}
0\\
.\\
.\\
.\\
1\\
\end{array}
\right)
\end{equation}
and is invariant under the action of a left $SU(N-1)$ subgroup and a diagonal
$U(1)$ group:
\begin{equation}
 {\displaystyle{z_0 }}\rightarrow e^{{\displaystyle{D_{N-1} \theta}}}\ \
 {\displaystyle{ z_0 }}\ \
e^{  {\displaystyle{i\sqrt{ {2 (N-1)\over N}}\ \ \theta}}},
\end{equation}
where $D_{N-1}$ is given by:
\begin{equation}
D_{N-1} = i\sqrt{\frac{2}{N\left(N-1\right)}}\left(\begin{array}{c}
1\\
.\\
.\\
.\\
1-N\\
\end{array}
\right).
\end{equation}
Action (\ref{action1}) thus describes the symmetry breaking pattern
$SU(N) \otimes U(1) \rightarrow SU(N-1) \otimes U(1)$.
It is a  NL$\sigma$ model defined on the homogeneous coset space $SU(N) \otimes
U(1)/SU(N-1)\otimes U(1)$ that
accounts for the physics of the resulting $2 N -1$ interacting
Goldstone modes.

Let us describe the symmetries associated with action (\ref{action1}) according
to the values taken by the coupling constants  $f_1$ and $f_2$. When $4 f_1 =
f_2$, action (\ref{action1}) becomes gauge invariant with respect
to the right $U(1)$ local group and one recovers the $CP^{N-1}$ model
relevant for non frustrated magnets$^{\cite{read3}}$.
In all other cases the gauge invariance
is broken as a consequence  of frustration on the microscopic spin system.
In the limit $f_2 \rightarrow \infty$,
the model becomes $O(2N)$ invariant as can be seen when parametrizing
the $z$ field in terms of a $2N$ component real vector field
${\pmb{$\phi$}}$.
The model thus interpolates smoothly, as one varies $f_2$, between the
two symmetric $CP^{N-1}$ and $O(2N)/O(2N-1)$  NL$\sigma$ models.

The partition function (\ref{fonctpart}) can be rewritten into another form
which enlightens the symmetry
properties and the role played by frustration in the model. As in the
$CP^{N-1}$
model, one can decouple the current-current interaction:
\begin{equation}
\omega^2_{\mu}=(z^{\dagger} \stackrel{\leftrightarrow}{\partial}_{\mu} z)^2
\label{courant}
\end{equation}
with help of a Lagrange gauge field $\cal{A}_{\mu}$$^{\cite{dadda}}$.
After some standard
manipulations, we obtain so a form that brings out more clearly gauge
invariance
properties:

\begin{equation}
Z= \int D{z^{\dagger}} Dz\ D{\cal A_{\mu}}\  \delta{\left(z^{\dagger}
z-{1\over 2f_1}\right)}
 \exp \left[ - \cal{S} \right]
\label{fonctpart2}
\end{equation}
with:
\begin{equation}
{\cal S} = \int d^{d} x \left[{\cal D}_{\mu} z^{\dagger}{\cal D}_{\mu} z
 + \Lambda^{d-2}\frac{M^2}{2} {\cal A_{\mu}}^{2} \right],
\label{actionmass}
\end{equation}
where ${\cal D}_{\mu} = {\partial}_{\mu} + i {\cal A_{\mu}}$ and
\begin{equation}
M^2 =  - \frac{1}{f_{1}}
\left( 1 - \frac{f_{2}}{4 f_{1}} \right).
\label{mass}
\end{equation}

In the absence of the  mass term $M^2$, i.e. for $4 f_{1}=f_{2}$,
action (\ref{actionmass}) is gauge invariant with respect
to the right $U(1)$ local group and  describes
the pure $CP^{N-1}$ model. When  $M^2 \ne 0$, i.e. $4 f_{1} \ne f_{2}$,
action (\ref{actionmass}) corresponds to a ``massive $CP^{N-1}$'' model.
We thus see that, in this parametrization, the effect of
frustration  is to give a mass to the gauge field ${\cal A_{\mu}}$.
As an immediate consequence we can already anticipate that
a new length scale has been generated by frustration. The
next sections will be largely devoted to discuss in great details
the physical and technical implications  of the existence of this new length
scale.

\section{WEAK COUPLING ANALYSIS}

\subsection{Introduction}

The renormalizability of NL$\sigma$ models defined on coset space $G/H$ around
two dimensions
 has been studied by Friedan$^{\cite{friedan}}$. He has shown that the RG
properties of a NL$\sigma$ model  only depend on the geometry of the coset
$G/H$
viewed as a metric manifold. In particular, the $\beta$ functions for the
coupling constants entering in the model are given, at two loop order, in terms
of the Ricci and Riemann tensors on the manifold $G/H$ so that renormalization
depends only on the {\sl local} properties of $G/H$ and is insensitive to its
global structure. It is worth stressing that this weak coupling expansion is,
in fact, a loop expansion ordered in powers of the temperature $T$. In
symmetric
models
such as $O(N)$ or $CP^{N-1}$ models, there is only one coupling which
identifies with the temperature. However, in non symmetric models such as
the massive $CP^{N-1}$ model, there are at least two coupling
constants. In this
case, the resulting expansion  is ordered in powers of an effective temperature
but is not perturbative
with respect to the ratio of the coupling constants. This is the reason why
the RG functions may
interpolate between fixed
points that lie at finite distance. With this respect, the low temperature
approach of the NL$\sigma$ model provides the natural
framework to study frustrated systems whose phase diagram exhibits several
fixed points. However, one has to keep in mind that the low temperature
expansion misses terms of order $e^{-1/T}$ that include
the global nature of the manifold - or of  the massive modes of the theory -
that may be
important for the physics$^{\cite{zinn}}$. We shall return to this point in the
next section
when dealing  with the ${1/N}$ expansion.

Starting from the partition function  (\ref{fonctpart}) we can take advantage
of the
 constraint of unit modulus to integrate out one degree of freedom of $z$.
To this end, let us parametrize $z$ as:
\begin{equation}
z=
\left(
\begin{array}{c}
{\pmb{$\pi$}}\\
e^{i\pi}\sqrt{1 - {{\pmb{$\pi$}}^{\dagger}{\pmb{$\pi$}}}}
\end{array}
\right),
\end{equation}
where ${\pmb{$\pi$}}$ is a $N-1$ component complex vector
field which, together
with the scalar field $\pi$, represent the $2N-1$ Goldstone modes.
With this parametrization, the partition function can be
rewritten in a more suggestive form:
\begin{equation}
Z=\int_{\vert \pi\vert\le 1} D{\pi} \
\exp \left[ -{{1\over 2}\int d^dx\ g_{ij}(\pi)\
\partial\pi^i\partial\pi^j} \right]
\label{toto2}
\end{equation}
which exhibits the geometrical nature of the theory. Here,
$\pi^i$ is a short notation for the $2N-1$ Goldstone modes.
The action $g_{ij}(\pi) \ \partial\pi^i\partial\pi^j$ represents the line
element on $G/H$ equipped with the metric $g_{ij}(\pi)$. Thus, in the absence
of the constraint $\vert \pi\vert\le 1$, action (\ref{toto2}) would
describe a free theory on the manifold $G/H$.
In the low temperature expansion this constraint is irrelevant at any finite
order of
perturbation theory  since it gives contributions
of order $e^{-1/T}$. Friedan has shown that the
$\beta$ functions which give the evolution of the metric with the scale
$\lambda$:
\begin{equation}
\beta_{ij}=\lambda{\partial g_{ij}\over \partial \lambda}\
\label{evolution}
\end{equation}
are given at two loop order by$^{\cite{friedan}}$:
\begin{equation}
\beta _{ij}(g)=   \epsilon g_{ij} -{1\over 2\pi} R_{ij} - {1\over8\pi^2}
R_{i}^{  pqr}R_{jpqr}
\label{beta}
\end{equation}
where $R_{ij}$ and $R_{jpqr}$ are respectively the Ricci and Riemann
tensors of the manifold $G/H$ equipped with the metric $g_{ij}$.
As we shall now see, these intrinsic properties can be formulated in the
language of group theory which enlightens the symmetry properties and
provides a powerful framework for practical computations.

\subsection{Vielbein basis formulation}

The main advantage of the geometrical interpretation of the RG properties
of NL$\sigma$ models is that the Goldstone modes $\pi^i$
(respectively the tangent vectors  $\partial_{\mu}\pi^i$) can be viewed
just as a particular coordinate frame on the manifold $G/H$ (respectively  on
the tangent space of $G/H$).
For practical calculations, it is extremely convenient to work,
in the tangent space, in the vielbein basis $\omega^I_\mu$ related to
the $\partial_{\mu} \pi^i$ via $\omega^I_\mu = \omega^I_i \partial_{\mu} \pi^i$
such that the vielbein basis metric $\eta_{IJ}=g_{ij}(\pi)\ \omega^i_I
\omega^j_J$ is free of any coordinate dependence. From the RG point of view,
the main advantage of the vielbein formulation is that the
geometrical quantities such as the Riemann tensor only depend on the Lie
algebras Lie$(G)$ and Lie$(H)$, i.e. on the structure constants
defined by the following commutation rules:
\begin{equation}
\left\{
\begin{array}{l}
[T_a,T_I] = {f_{aI}}^JT_J \ ,\\
{[T_a,T_b] = {f_{ab}}^cT_c}\ ,\\
{[T_I,T_J] = {f_{IJ}}^KT_K + {f_{IJ}}^aT_a}\ ,
\end{array}
\label{fstructure}
\right.
\end{equation}
where $T_a\in$ Lie$(H)$ and $T_I\in$ Lie$(G)$-Lie$(H)$ are
normalized according to  ${\hbox {Tr}}\ T_iT_j=-2\delta_{ij}$.
We shall now see how this works in the massive $CP^{N-1}$ model.

Let us parametrize the coset $SU(N) \otimes U(1)/SU(N-1)\otimes U(1)$ by
choosing one unique element, i.e. fixing the gauge:
\begin{equation}
L(\pi^i)= \left(\begin{array}{c}
\sqrt{1-{\pmb{$\pi$}} {\pmb{$\pi$}}^{\dagger}}\\
- e^{-i\pi} \ \ {\pmb{$\pi$}}^{\dagger}
\end{array}
\right.
\left.
\begin{array}{c}
{\pmb{$\pi$}}\\
e^{i \pi}\sqrt{1- {\pmb{$\pi$}}^{\dagger} {\pmb{$\pi$}}}
\end{array}
\right).
\end{equation}
This matrix is the same as for the coset space
$SU(N) /SU(N-1)$. This is the consequence of the fact that
the coset space $G\otimes X/H\otimes X$ - where $X$ is the maximal subgroup
of $G$ that commutes with $H$ - and $G/H$ are topologically equivalent. In the
following,
we shall work directly in the coset $SU(N)/SU(N-1)$ keeping in mind
that we look for an action which has $SU(N)\otimes U(1)$ as symmetry
(isometry) group.

The quantity
$L^{-1}\partial L$ belongs to Lie$(G)$ of $G=SU(N)$
and we have:
\begin{equation}
L^{-1}\partial L=(L^{-1}\partial L)_{G-H}+(L^{-1}\partial L)_{H}
\label{ldl}
\end{equation}
where $(L^{-1} \partial L)_{H}$ is in  Lie$(H)$ of $H=SU(N-1)$.
One can now express the action (\ref{action1}) in terms of the currents
(\ref{ldl}) as:
\begin{equation}
{\cal S}=-{1\over 2}\int d^dx\  {\hbox{Tr}}[( L^{-1} \partial L)_{G-H}]^2\
\label{actiona}
\end{equation}
which is the general expression of NL$\sigma$
models. It depends only on the currents in Lie$(G)$-Lie$(H)$, the tangent
space of $G/H$.
This reflects the gauge invariance under the $H$ group, i.e.
the arbitrary in the choice of the ground state.
Since $L^{-1}\partial L$ belongs to Lie$(G)$, eq.(\ref{ldl}) can be
rewritten as:
\begin{equation}
L^{-1}\partial_\mu L = \omega_\mu^I T_I +\Omega_\mu^a T_a\ ,
\label{viel}
\end{equation}
where
$\omega_\mu^I$ and $\Omega_\mu^a$ are
respectively the vielbein and the connection in the tangent space
of $G/H$. They are given by:
\begin{equation}
\left\{
\begin{array}{lll}
\omega_\mu^I &=&  -\displaystyle {1\over 2}
{\hbox{Tr}}\left[ L^{-1}\partial_\mu L \ \ T_I\right]\\
\\
\Omega_\mu^a &=&  -\displaystyle {1\over 2} {\hbox{Tr}}\left[
L^{-1}\partial_\mu L \ \ T_a\right].
\end{array}
\label{current}
\right.
\end{equation}
Among all these currents, the $U(1)$ one plays a
crucial role. It is defined in terms of the $z$ field as:
\begin{equation}
\begin{array}{ll}
\omega_\mu^X & =  - {\displaystyle{\frac{1}{2}}} {\hbox{Tr}}\left[
L^{-1}\partial_\mu L
\ \ D_{N-1}\right] = {\displaystyle{ {i\over 4} \sqrt{ {2 N\over N-1}}}}\ \
\left( z^{\dagger} \stackrel{\leftrightarrow}{\partial}_{\mu} z \right)\
\end{array}
\label{currentU(1)}
\end{equation}
while the currents
$\omega_\mu^A$  in Lie$(SU(N))$-Lie$(SU(N-1))$-Lie$(U(1))$ verify:
\begin{equation}
\sum_A {(\omega_\mu^A)}^2 = \left[
\partial_{\mu} z^{\dagger} \partial_{\mu} z
+ {1\over 4}
\left( z^{\dagger} \stackrel{\leftrightarrow}{\partial}_{\mu} z \right)^2
\right
].
\label{ter}
\end{equation}
We can now write action (\ref{action1}) in terms of these currents:
\begin{equation}
{\cal S} = {1\over 2} \int d^dx\ \left[ \eta_1
\sum_A {(\omega_\mu^A)}^2 + \eta_2 {(\omega_\mu^X)}^2 \right]\
\label{actioncourrant}
\end{equation}
which defines the vielbein basis metric $\eta_{IJ}$. Indeed we have:
\begin{equation}
S={1\over 2}\int d^dx\ \eta_{IJ}\ \omega^I_\mu \omega^J_\mu\ ,
\label{actiontangent}
\end{equation}
where
\begin{equation}
\eta_{IJ}=-{1\over
2}\left[\eta_1{\hbox{Tr}}(T^IT^J)+(\eta_2-\eta_1)\delta_{I\alpha}
\delta_{J\beta}{\hbox{Tr}}(T^\alpha T^\beta)\right],
\label{eta}
\end{equation}
with:
\begin{equation}
\left\{
\begin{array}{lll}
\eta_1 & = {\displaystyle{{1\over f_1}}} \\
\eta_2 & = 2 \ \ {\displaystyle{
 {N-1\over Nf_1}}} \left( 1 - {\displaystyle{{4f_1\over f_2}}} \right)\ .
\end{array}
\label{etaf}
\right.
\end{equation}

We recall that in our case $T_I\in$ Lie$(SU(N))$-Lie$(SU(N-1))$,
$T_\alpha\in$ Lie$(U(1))$ and
$T_a\in$ Lie$(SU(N-1))$ and that the corresponding algebra is given in
(\ref{fstructure}). As already pointed out,
whereas non frustrated models are defined with one coupling constant,
the frustrated case is characterized by two independent
coupling constants $ \eta_1$ and $ \eta_2$.

The crucial advantage of the vielbein basis formulation is that the
Riemann tensor only depends on the structure constants $f_{ij}^{\;\; k}$ of
Lie$(G)$ so that the $\beta$ functions:
\begin{equation}
\beta_{IJ}=\lambda{\partial \eta_{IJ}\over \partial \lambda}\ ,
\end{equation}
writes at two loop order:
\begin{equation}
\beta _{IJ}(\eta)=\epsilon \eta_{IJ} - {1\over 2\pi}R_{IJ} -
{1\over8 \pi^2} \ R_{I}^{PQR}\ R_{JPQR}
\label{betaij}
\end{equation}
where the Riemann tensor writes$^{\cite{aza4}}$:
\begin{eqnarray}
R_{IJKL} \hspace{-0.3cm}&=&\hspace{-0.3cm}f{_{IJ}}^af_{aKL} + {1\over
2}{f_{IJ}}^M\left( f_{MKL}+f_{LMK}-f_{KLM}\right)\nonumber\\
\hspace{-0.3cm}& &\hspace{-0.3cm}+{1\over4}\left(f_{IKM}
+f_{MIK}-f_{KMI}\right)\left(
{{f_J}^M}_L + {f_{LJ}}^M -{f^M}_{LJ}\right)\nonumber\\
 \hspace{-0.3cm}& &\hspace{-0.3cm}-{1\over4}\left(f_{JKM}
+f_{MJK}-f_{KMJ}\right)\left(
{{f_I}^M}_L + {f_{LI}}^M -{f^M}_{LI}\right).
\label{riemann}
\end{eqnarray}
The indices $a$ and $\{I,J\dots\}$ refer to $H$ and $G-H$ respectively. The
$G-H$ indices are raised and lowered by means of $\eta ^{IJ}$ and
$\eta _{IJ}$ and repeated indices are summed over.

\subsection{Results for the ${\bf SU(N) \otimes U(1)/SU(N-1)\otimes U(1)}$
model}

We are now in a position to study the RG properties of the massive
$CP^{N-1}$ model. The computation of the Riemann tensor (eq.(\ref{riemann}))
leads to
the following  $\beta$ functions:
\begin{equation}
\left\{
\begin{array}{lll}
{\displaystyle{\lambda
{\partial\eta_1\over\partial
\lambda}}}&\hspace{-0.3cm}=&\hspace{-0.3cm}\epsilon
 \eta_1 + {\displaystyle{\beta_{\eta_1}}}\\
&&\\
{\displaystyle{\lambda\frac{\partial{\bar \eta_2}}{\partial \lambda}}} &
\hspace{-0.3cm}=&\hspace{-0.3cm}\epsilon
 {\bar \eta_2} + {\displaystyle{ \beta_{{\bar \eta_2}}}}
\end{array}
\right.
\label{eqrecursion}
\end{equation}
with:
\begin{equation}
\left\{
\begin{array}{lll}
{\displaystyle{\beta_{\eta_1}}}&\hspace{-0.3cm}=&\hspace{-0.3cm}-
{\displaystyle{
 \frac{1}{\pi}}}\left(N -{\displaystyle{
\frac{{\bar
\eta_2}}{\eta_1}}}\right)-\displaystyle{\frac{1}{\eta_1}{\frac{1}{2\pi^2}}}\left(
4 N - 6 \ N\  {\displaystyle{\frac{{\bar \eta_2}}{\eta_1}}}
+{\displaystyle{\ (3N-1)\left(\frac{{\bar \eta_2}}{\eta_1}\right)^2}} \right)\\
&&\\
{\displaystyle{ \beta_{{\bar \eta_2}}}}  & \hspace{-0.3cm}=&\hspace{-0.3cm} -
{\displaystyle{\frac{N-1}{\pi}
\left(\frac{{\bar \eta_2}}{\eta_1}\right)^2 -{\displaystyle{\frac{1}{\eta_1}}}
\frac{N-1}{2 \pi^2}\left(\frac{{\bar \eta_2}}{\eta_1}\right)^3}},
\end{array}
\label{betaeta}
\right.
\end{equation}
where ${\bar \eta_2} = {\displaystyle{\frac{N}{2(N-1)} \eta_2}}$.

In eq.(\ref{eqrecursion}), the one and two loop order contributions to
the
$\beta$ functions are ordered in powers of
${\displaystyle{\frac{1}{\eta_1} = T}}$, the effective temperature of the
model.
The long distance, low energy, physics is determined by the behavior of the RG
flow in the $\lambda \rightarrow \infty$ limit while critical behavior is
associated
with the existence of  UV stable fixed points of
(\ref{eqrecursion}). To exhibit the low $T$ properties, it is convenient
to make the following change of variables:
\begin{equation}
\left\{
\begin{array}{lll}
T & = {\displaystyle{{1\over \eta_1}}} = f_1\\
x & = 1- {\displaystyle{ \frac{{\bar \eta_2}}{\eta_1}}} =
{\displaystyle{{4f_1\over f_2}}}\ .
\end{array}
\right.
\label{tx}
\end{equation}
In terms of these new coupling constants, eq.(\ref{eqrecursion}) reads:
\begin{equation}
\left\{
\begin{array}{lll}
{\displaystyle{\lambda
{\partial T\over\partial \lambda}}}&\hspace{-0.3cm}=&\hspace{-0.3cm}-\epsilon \
 T + \beta_T\\
&&\\
{\displaystyle{\lambda\frac{\partial x}{\partial \lambda}}} &
\hspace{-0.3cm}=&\hspace{-0.3cm}\beta_x
\end{array}
\right.
\label{eqrecursiontx}
\end{equation}
where:
\begin{equation}
\left\{
\begin{array}{lll}
{\displaystyle{\beta_T }}&\hspace{-0.3cm}=&\hspace{-0.3cm}
\displaystyle \frac{T^2}{\pi}\left(N -1 +
x\right) + \displaystyle \frac{T^3}{2\pi^2}\left(N-1+2x+(3N-1)\ x^2\right)\\
&&\\
{\displaystyle{\beta_x}} & \hspace{-0.3cm}=&\hspace{-0.3cm}
- \displaystyle N\ \frac{T}{\pi}\ x\ (1-x) - \displaystyle N\
\frac{T^2}{\pi^2}\ x\ (1-x^2)\ .
\end{array}
\right.
\label{betatx}
\end{equation}

As usually, $T$ is the relevant coupling constant that orders perturbation
theory.
The variable $x$ measures the anisotropy of the model: as $x$ varies
one interpolates between the pure $CP^{N-1}$ model, when $x=1$, and the pure
$O(2N)$ model, when $x=0$. In the following, we shall study separately
the $d=2$ and $d > 2$ cases.

\subsubsection{Two dimensional case}

When $\epsilon =0$, eq.(\ref{eqrecursiontx}) exhibits a line of UV
fixed points at $T=0$ for any values of $x$.
However they are not all true fixed points in the sense of limit of RG
trajectories. To get the complete RG properties one has to solve the flow
resulting from eq.(\ref{eqrecursiontx}).
In the following, we restrict ourselves to one loop accuracy
since inclusion of two loops
terms does not modify qualitatively our conclusions.

The RG trajectories are parametrized by the RG invariant
$K_N$, given at one loop by:
\begin{equation}
K_N= {\displaystyle{ \frac{x^{1-1/N}}{1-x} T}},
\label{flowinvariant}
\end{equation}
in terms of which the solutions of eq.(\ref{eqrecursiontx}) are given by
the implicit equation:
\begin{equation}
\left\{
\begin{array}{lll}
 K_N\ {\ln}\lambda&=& A_N\left(x(1)\right)-A_N\left(x(\lambda)\right)\\
&&\\
T(\lambda)&=& K_N {\displaystyle{ \frac{1-x(\lambda)}{x(\lambda)^{1-1/N}}}}
\end{array}
\right.
\label{implicit}
\end{equation}
with initial conditions  $x(1)$ and $T(1)$ satisfying eq.(\ref{flowinvariant}).
The function $A_N(x)$ is given in terms of the hypergeometric function $_2F_1$:
\begin{equation}
A_N(x)= {\displaystyle{ \frac{\pi}{N-1} x^{1-1/N} \\ {_2F_1}\left(2,
1-\frac{1}{N}; 2-\frac{1}{N}; x\right)}}\ .
\label{ter2}
\end{equation}
We show in Fig.1 the corresponding flow diagram.
%\vspace{7cm}

In the UV regime, i.e. when $\lambda\to 0$, all the flow lines
with $K_N  \neq 0$ reach the $CP^{N-1}$ fixed point
at $x=1$ and $T=0$ which thus governs the short distance
behavior of  models with $x \neq  0$. On the invariant line $K_N=0$,
corresponding to the $O(2N)$ model, the flow is governed
in the UV by the $O(2N)$ fixed point at $x=0$ and $T=0$.
In the infrared limit, when $\lambda$ increases, $x(\lambda)$ decreases to
its asymptotic value at $x=0$. This means that all models with $x(1) \neq  1$
are asymptotically equivalent at long distance to the $O(2N)$ model.
We thus conclude that, while a {\sl priori} we have a line
of fixed points at $T=0$, only two of them determine the RG properties.
They are the $CP^{N-1}$ and $O(2N)$ fixed points
that govern the short and long distance behaviors respectively.
The presence of these two competing fixed points is a characteristic
feature of  frustration  where fluctuations of different
nature  compete. We are thus led to expect that apart from
the correlation length $\xi$, there should exist a new scale $\Xi$,
characterizing the cross-over between short and long wavelength
associated with the $CP^{N-1}$ and $O(2N)$ fixed points respectively. As a
consequence,
the scaling behavior of correlation functions will
depend, in a non trivial way, on the relative magnitude of $\xi$,
$\Xi$ and $\Lambda^{-1}$. To see this, we have to
go further in the flow analysis.

First of all, in order to perturbation theory remains valid
one must have $T(\lambda) \ll 1$.
Given initial conditions on the flow $(x(1),T(1))$ at scale $\Lambda^{-1}$,
this implies that $\lambda$ has to be much smaller
than $\lambda_0$ defined by:
\begin{equation}
A_N\left(x(1)\right)=K_N\ {\ln}\lambda_0\ .
\end{equation}
This defines the correlation length $\xi = \lambda_0 \Lambda^{-1}$ of the
model:
\begin{equation}
\xi = \Lambda^{-1} {\displaystyle{ \exp \left( \frac{A_N\left( x(1)
\right)(1-x(1))}{x(1)^{1-1/N} T(1)}\right)}},
\label{correl}
\end{equation}
which follows the standard RG equation:
\begin{equation}
{\displaystyle{\left( \Lambda \frac{\partial}{\partial \Lambda} + \beta_x
\frac{\partial}{\partial x} + \beta_T \frac{\partial}{\partial T} \right) \xi =
0}}\ .
\label{rginv}
\end{equation}
The general solution of eq.(\ref{rginv}) writes in fact $\Xi = \Phi(K_N)\ \xi$
where $\Phi(K_N)$ is an arbitrary function of the flow invariant $K_N$.
This means that in a model with several coupling constants there may be others
physical scales necessary to characterize the complete scaling behavior. We
shall indeed see that if $\lambda_0$ defines the critical regime
through the relation $1\ll\lambda \ll \lambda_0$ it does not specify
which fixed point governs the scaling behavior of the correlation functions.
Let us now precise this point by defining three regimes according
to the initial conditions on the flow:

$\bullet$ Given initial conditions very close to the $CP^{N-1}$ fixed point
with $x(1) \sim 1$, we define the region governed by the
$CP^{N-1}$ fixed point to be such that $x(\lambda)$ is still close to 1.
Since in this regime $A_N(x) \sim {\displaystyle{\frac{\pi}{N(1-x)}}}$,
one finds that $\lambda$ has to be much smaller than
\begin{equation}
{\displaystyle{\lambda_1}}= {\displaystyle{\lambda_0}}
 \exp\left( {\displaystyle{ -\frac{\pi}{NK_N}}} \right).
\label{l1}
\end{equation}
We so obtain a new RG invariant scale $\Xi = \lambda_1 \Lambda^{-1}$,
solution of eq.(\ref{rginv}), that defines the region governed by
the $CP^{N-1}$ fixed point: $1 \ll \lambda \ll \lambda_1$ or, in terms of the
momentum
$p$, $\Xi^{-1}\ll p \ll \Lambda$. Notice that in this regime one has
$1 - x(\lambda) \sim T(\lambda)/K_N$.

$\bullet$ Independently of the initial conditions $x(1)$, the $O(2N)$ regime
is obtained for values of $\lambda \ll \lambda_0$ such that $x(\lambda) \sim
0$. One sees that it is sufficient to have
$\lambda \gg \lambda_1$ since $A_N(x) \sim {\displaystyle{\frac{\pi}{N-1}
x^{1-1/N}}}$ when $x \rightarrow 0$. The latter condition
not only implies $K_N \ll 1$ but also defines the region governed by
the $O(2N)$ fixed point: $\lambda_1 \ll \lambda \ll \lambda_0$ or $ \xi^{-1}\ll
p \ll \Xi^{-1}$.

$\bullet$ Finally, when $K_N \ll 1$ and $x(1) \sim 1$, one
has $\xi \gg \Xi$ so that
the $O(2N)$ and $CP^{N-1}$ regimes can be observed
when $\xi^{-1} \ll p \ll \Xi^{-1}$ and $\Xi^{-1} \ll p \ll \Lambda$
respectively.

Let us resume our findings. We first found that the
perturbation theory remains valid if one stays
in a regime $\lambda \ll \Lambda\xi$ so that $T(\lambda) \ll 1$.
In symmetric NL$\sigma$ models, such as when $x=1$ or $x=0$, the
correlation functions display scaling behavior
with logarithmic corrections that are governed by the unique non trivial fixed
point
of the theory. In non symmetric NL$\sigma$ models, i.e. $x(1) \ne 0,1$,
even when $\lambda \ll \Lambda\xi$, there remains a non trivial scale
in the theory. In both regimes $\xi^{-1} \ll p \ll \Xi^{-1}$
and $\Xi^{-1} \ll p \ll \Lambda$, i.e. when one lies near
the $O(2N)$ or $CP^{N-1}$ fixed points, the field theoretical approach
to the RG is expected to give the correct
asymptotic behavior of the correlations functions.

The previous discussion  can
be formulated in terms of the gauge field formulation (\ref{actionmass})
of the theory. This is most easily done in the limit $N \rightarrow \infty$,
where $\Xi$ takes the nice form:
\begin{equation}
{\Xi} =\xi e^{\displaystyle{-\pi M^2\over N}}
\label{ximass}
\end{equation}
where $M^2$ is the gauge field mass (see eq.(\ref{mass})). We see
that $\xi$ being fixed, the length ${\Xi}$ is associated
with the energy scale where the effects of the $U(1)$ current-current
interaction
become strong. We shall return  to this point in more details when studying the
$1/N$ expansion. Let us just finally observe that the status of the $U(1)$
current-current
interaction of eq.(\ref{ter}) changes between the $CP^{N-1}$
and $O(2N)$ fixed points. While it is marginally relevant near the former, it
becomes marginally irrelevant near the latter fixed point and
is therefore a ``dangerous irrelevant'' variable.

All the previous discussion is not qualitatively modified by inclusion of two
loop order terms in the $\beta$ functions. For completeness,
we now give the expression of the correlation length $\xi$ at two loop order:
\begin{equation}
{\displaystyle{ \xi= \xi_{1-loop} \ G(x) \left( \psi_0 \
T^{{\displaystyle{ \frac{1}{2(N-1)}}}} \  +
\psi_1 \ T^{{\displaystyle{\frac{2}{N}}}} \  {\displaystyle{\frac{ x^{
{\displaystyle{ \frac{3N-4}{2N^2}}  }}}{(1-x)^{
{\displaystyle{\frac{3N-4}{2N(N-1)} }} }}}} \right)}}
\label{twoloopcorr}
\end{equation}
where $\xi_{1-loop}$ is the one loop correlation length given
by eq.(\ref{correl}), $\Psi_0$ and $\Psi_1$ being constants and $G(x)$ being a
pure function of $x$ given by:
\begin{equation}
G(x) =\exp{{\displaystyle{\int dx }} \left( {\displaystyle{  \frac{
{_2F_1}\left(2, 1-\frac{1}{N}; 2-\frac{1}{N}; x\right)(1-x)(3x+1)}{2Nx} }}
- {\displaystyle{\frac{  N-1 +(2N-3) x}{2N(N-1)x(1-x)}}}  \right)}
\label{Tehta}
\end{equation}
obtained by solving eq.(\ref{rginv}).

In the very neighborhood of $O(2N)$ fixed point, eq.(\ref{twoloopcorr})
becomes to leading order in $x$:
\begin{equation}
\xi = \xi_{O(2N)} \ \left( 1 + \frac{x}{4N^2} \right) \ \exp \
\left({\displaystyle{\frac{-x\pi}{T(2N-1)(N-1)}}}\right),
\label{xio2N}
\end{equation}
while in the neighborhood of the $CP^{N-1}$ fixed point,
where  ${\displaystyle{1-x= \frac{T}{K_N}}}+O(T^2)$, one finds
to leading order in $1-x$ and $T$:
\begin{equation}
\xi = \xi_{CP^{N-1}} \exp \
\left({\displaystyle{-\frac{\left(1-x\right)\pi}{N^2T}}{\ln}
\frac{T}{2\pi}}\right)
\label{xiCPN}
\end{equation}
where $\xi_{O(2N)}$ and $\xi_{CP^{N-1}}$ are the two loop
correlation lengths of the $O(2N)$ and $CP^{N-1}$
models$^{\cite{brezin,hikami1}}$:
\begin{equation}
\left\{
\begin{array}{lll}
{\displaystyle{\xi_{O(2N)}=C_{1\xi}\  {\Lambda}^{-1}\  T^{{\displaystyle{
\frac{1}{2(N-1)}}}}\  \exp\left({\displaystyle{{\pi\over {(N-1)T}}}}\right)}}
\\
{\displaystyle{\xi_{CP^{N-1}}=C_{2\xi}\  {\Lambda}^{-1}\  T^{{\displaystyle{
\frac{2}{N}}}}\  \exp  \left({\displaystyle{{\pi\over {NT}}}}\right)}}
\end{array}
\label{fstructure1}
\right.
\end{equation}
where $C_{1\xi}$ and $C_{2\xi}$ are constants that depend on the
regularization scheme.

\subsubsection{Above two dimensions}

In dimension $d > 2$, there is a phase transition, as one varies $T$, from
an ordered phase, with the breaking of the $SU(N) \otimes U(1)$ symmetry, to a
disordered phase. Indeed, eq.(\ref{eqrecursiontx}) admits, apart
from the trivial $T=0$ fixed point, two non trivial fixed points with
$(\displaystyle{T_{CP^{N-1}}^{*} ={\pi\epsilon\over N}-{2\pi\epsilon^2\over
N^2}, x^{*} =1)}$
and $\displaystyle{(T_{O(2N)}^{*} ={\pi\epsilon\over N-1}-{\pi\epsilon^2\over
2(N-1)^2}, x^{*} =0)}$ which are respectively $CP^{N-1}$ and $O(2N)$
symmetric. In the infrared regime, the latter has one direction of instability
and thus governs the long distance behavior while the former
has two unstable directions and governs the short distance
behavior. On the critical line connecting the two non trivial
fixed points, the flow drives the system towards the $O(2N)$ fixed
point which is thus responsible of the phase transition of
frustrated models above two dimensions (see Fig.2).
The symmetry is increased at the stable IR fixed point
for all values of $N$. In particular when $N=2$, we find the $O(4)/O(3)$
universality class for frustrated Heisenberg spin systems in agreement
with  previous studies of the $O(N)\otimes O(2)/O(N-2)\otimes O(2)$
NL$\sigma$ model with $N=3$$^{\cite{aza4}}$.

As in two dimensions, there exists a specific
scale $\Xi_{\epsilon} =\lambda_{1\epsilon} \Lambda^{-1}$
associated with the cross-over between the $O(2N)$
and $CP^{N-1}$ behavior. The scale $\Xi_{\epsilon}$ can be again obtained by a
careful
analysis of the RG flow. As previously, we shall restrict ourselves to one
loop accuracy.

The one loop RG invariant flow takes now the form:

\begin{equation}
K_N= {\displaystyle{ \frac{x^{1-1/N} }{1-x} (T-T_c(x))}},
\label{flowinvariant2}
\end{equation}
with:
\begin{equation}
T_c(x)={\displaystyle{{\pi \epsilon\over N-1}{(1-x)}\;{_2F_1}\left(2,
1-\frac{1}{N}; 2-\frac{1}{N}; x\right)}}\ .
\label{tcritic}
\end{equation}
The critical line $T_c(x)$ is obtained when $K_N = 0$, and $K_N >0$ label
the lines in the disordered phase. With help of these quantities,
the solutions of eq.(\ref{eqrecursiontx}) are given by:
\begin{equation}
\left\{
\begin{array}{lll}
\lambda^{-\epsilon}&=&{\displaystyle{{{\epsilon A_N}(x(\lambda))+K_N \over
{\epsilon A_N}(x(1))+ K_N}}}\\
\\
T(\lambda)&=&{\displaystyle{\frac{1-x(\lambda)}{x(\lambda)^{1-1/N}}\left(
K_N+{{\epsilon A_N}(x(\lambda))}\right)}}
\end{array}
\right.
\label{solution}
\end{equation}
where $A_N$ is given by eq.(\ref{ter2}).
In the symmetric phase, when $K_N >0$, the analysis of the RG flow follows
straightforwardly from that of $d=2$. We obtain for the
correlation length and the cross-over scale:
\begin{equation}
\left\{
\begin{array}{lll}
\xi &=&\Lambda^{-1} {\displaystyle{  \left( 1 + \epsilon { A_N(x(1)) \over K_N}
\right)^{1/\epsilon}    }}\\
\\
\Xi_{\epsilon} &=&\xi {\displaystyle{  \left(1 + {\epsilon \pi \over N K_N}
\right)^{-1/\epsilon}}}
\end{array}
\right.
\label{xiXisup2}
\end{equation}
which define, exactly as in $d=2$, the regimes of energy governed by either
the $CP^{N-1}$ or the $O(2N)$ fixed point. On the critical line, i.e.
when $K_N  \rightarrow 0$, $\xi$ diverges but $\Xi$ has a finite limit:
$\Xi_{c} = \Lambda^{-1} {\displaystyle{\left(\frac{N
A_N(x(1))}{\pi}\right)^{{1/ \epsilon}}}}$.
This means that even on the critical line, the $O(2N)$ critical behavior
of frustrated  systems can only be seen asymptotically
in the regime $p\ll \Xi_{c}^{-1}$.
This fact may be of importance when studying frustrated spin systems
on the lattice where one has to work with lattice sizes $L \gg \Xi_{c}$ to be
able to observe $O(2N)$ universal scaling. In the limit $N \rightarrow \infty$,
$\Xi_c$ can be expressed in terms of $M^2$ of eq.(\ref{mass}) as:
\begin{equation}
\Xi_{c} = \Lambda^{-1} \left({\displaystyle{ {N \over \pi \epsilon M^2}}}
\right)^{1/\epsilon}.
\label{Ximasseps}
\end{equation}

Let us finally give the expressions of the critical exponents:

$\bullet$ At the $O(2N)$ fixed point, $x$ is an irrelevant variable and scales
to zero as:
\begin{equation}
x(\lambda) \sim \left( {\lambda \over \lambda_{1\epsilon}}\right)^{-
\phi_{O(2N)}}, \ \
 \lambda \gg \lambda_{1\epsilon}\ .
\label{scalo2n}
\end{equation}
This defines the cross-over exponent $\phi_{O(2N)}$
which is obtained, at two loop, from eq.(\ref{betatx}):
\begin{equation}
\phi_{O(2N)} = \frac{N}{N-1}\epsilon + \frac{N}{2\left(N-1\right)^2}\epsilon^2
+ O\left(\epsilon^3\right)\ .
\label{crossexp}
\end{equation}
The exponent $\nu_{O(2N)}$ itself is, of course, given by that of the $O(2N)$
model$^{\cite{brezin1}}$:
\begin{equation}
{\displaystyle{\nu^{-1}_{O(2N)}=\epsilon+{\epsilon^2\over
2N-2}+O(\epsilon^3)}}\ .
\label{nuo2n}
\end{equation}

$\bullet$ Near the $CP^{N-1}$ fixed point, $T$ and $x$ are relevant.
While the scaling exponent is given in the $T$ direction
by$^{\cite{hikami1}}$:
\begin{equation}
{\displaystyle{\nu^{-1}_{CP^{N-1}}=\epsilon+{2\over
N}\epsilon^2+O(\epsilon^3)}}\ ,
\label{nucpn}
\end{equation}
we find that $x(\lambda)$ scales trivially:
\begin{equation}
x(\lambda) \sim {\displaystyle{1 - \left({\lambda \over
\lambda_{1\epsilon}}\right)^{\epsilon}}}, \ \  \lambda \ll \lambda_{1\epsilon}
\label{scalcpn}
\end{equation}
so that $\phi_{CP^{N-1}}=\epsilon$\ .
This property holds at all orders
of perturbation theory. This is easily seen on eq.(\ref{betaeta})
where one can observe that the $\beta$ function for ${\bar \eta_2}$
orders with increasing powers of $({\bar \eta_2}/{\eta_1})^p$, $p\ge  2$,
so that all its partial derivatives with respect to ${\bar\eta_2}$
taken at ${\bar \eta_2} = 0$ vanishe.
Hence the scaling of ${\bar \eta_2}$ follows simply from
dimensional analysis. This means that ${\bar \eta_2}$ does not renormalize
at sufficiently high energy and that the $U(1)$ current
has no anomalous dimension at the $CP^{N-1}$ fixed point, a result
which is completely consistent with the $U(1)$ gauge invariance.

\subsection{Concluding remarks}

In this section, we have studied the massive $CP^{N-1}$ model by means of a low
temperature expansion around two dimensions. We have expressed the RG
properties of this model in terms of local, geometrical, quantities associated
with the manifold $SU(N) \otimes U(1)/ SU(N-1) \otimes U(1)$ viewed as a metric
space. We have found that the $\beta$ functions and
the mass gap, calculated up to two loop order, interpolate smoothly between
those of  $CP^{N-1}$ and $O(2N)$ models as one varies $x$. In addition,  a
careful analysis of the RG flow have revealed the existence of a new length
scale
$\Xi$, characterizing the cross-over from short distance behavior, governed
by the $CP^{N-1}$ fixed point, to the long distance one governed by the $O(2N)$
fixed point. The appearance of this length results from the competition between
fluctuations of different nature and is reminiscent of the
frustration of the lattice spin system. An important point is that, contrarily
to the correlation length
 $\xi$, the length $\Xi$ is not related
to a singular behavior of the RG flow so that perturbation theory is a priori
well defined in all the regime $ \xi^{-1}\ll p \ll \Lambda$ independently
of the relative value of $p$ and $\Xi$, or equivalently,
for any value of $x$ between 0 and 1. Indeed, the low temperature expansion
which is at the basis of the NL$\sigma$ model analysis is a loop expansion
ordered in powers of $T$ and $\epsilon$ and is insensitive to  the value of
$x$.

It is however worth to stress that all our results rely precisely on the
perturbative
renormalizability of the model in this
double expansion in $T$ and $\epsilon$.  Under this hypothesis, we expect that,
at sufficiently low energy, i.e. when $p\ll\Lambda$, the physics of the spin
system is well described in terms
of  only {\sl two} coupling constants $x$ and $T$ and is therefore universal.
In this regime, the high energy degrees of freedom have decoupled from the low
energy ones and  all the remaining irrelevant operators give
subdominant
contributions to the scaling of physical quantities.
The question that naturally arises is whether or not this universality property
holds beyond perturbation theory and more precisely, to what extent can we
control the effects of irrelevant operators beyond the double expansion in
$\epsilon$ and $T$. Indeed, one  possible source of failure of perturbation
theory in NL$\sigma$ models is the neglected $O(e^{-1/T})$ terms that take into
account of the global nature of the manifold.
The  best that one can do is to  check the consistency of the low $T$ expansion
against other perturbatives approaches such as the $1/N$ expansion. It is the
purpose of the next section to investigate the properties of the massive
$CP^{N-1}$ model through an expansion in the number of components of the fields
and to study some aspects of the relationship between the large  $N$ and weak
coupling analysis of this model.

\section{THE LARGE N APPROACH.}

The ${1/N}$ expansion  is a powerful tool on several accounts. On the one hand,
it
reveals by
itself non perturbative informations about the model: mass generation,
existence of a phase transition, etc. On the other hand, it provides a test of
consistency of perturbative approaches, like weak coupling analysis. However,
up to now, because of their strong analogies with gauge field theories in four
dimensions and their simplicity,  only  symmetric models like
$CP^{N-1}$$^{\cite{dadda,witten,cant,arefeva,campos7}}$ or
$O(2N)$$^{\cite{brezin1,bardeen1,rosenstein2,flyvbjerg}}$ NL$\sigma$ models -
with sometimes their supersymmetric counterpart$^{\cite{davis}}$ - and
Gross-Neveu model$^{\cite{grossneveu,davis}}$ in two dimensions have been
subject
to a detailed analysis.
In particular, it is just for the two last ones that the precise relationship
between the large  $N$ and weak coupling analysis has really been understood
beyond leading order$^{\cite{rim}}$. Such a relationship is still lacking in
the case
of more involved models like the massive $CP^{N-1}$ model .

As already emphasized, the main feature of this  model  is
that the current-current operator $(z^{\dagger}
\stackrel{\leftrightarrow}{\partial}_{\mu} z)^2$
which is relevant at the $CP^{N-1}$ fixed point  becomes irrelevant at the
$O(2N)$ one.
In the double expansion in $T$ and $\epsilon$ which is at the
basis of the low
 temperature approach of the NL$\sigma$ model around two dimensions, such a
term stays
 marginal and  is consequently renormalizable. This is why the low $T$
expansion allows to
 interpolate between the two fixed points of the theory. However this situation
is
 particular to the weak coupling expansion and one does not expect that this
will be the
 case in  other  perturbative scheme. Indeed, we shall see that the standard
$1/N$
expansion is
unable to describe the whole phase diagram of the massive $CP^{N-1}$ model.
The reason for this is that the cross-over scale that emerged from  the RG flow
analysis of the weak coupling expansion enters in the $1/N$ expansion in such
a way that it provides a boundary to the domain of validity of the standard
large $N$ perturbative approach. We shall show however that by carefully taking
into account of
the  presence of this new length scale, it is possible to capture the correct
renormalization of the parameters of the massive  $CP^{N-1}$ model in the large
$N$ analysis at least around each of the two fixed points of the model.

Let us first explicit
the $N$ dependence of the coupling constants $f_1$ and $f_2$. The action
(\ref{action1})
writes now:
\begin{equation}
{\cal S} = \int d^{d} x \left[ \Lambda^{d-2}\frac{N}{2f_{1}}\partial_{\mu}
z^{\dagger} \partial_{\mu} z
+ \Lambda^{d-2}\frac{N}{2f_{2}} \left( z^{\dagger}
\stackrel{\leftrightarrow}{\partial}_{\mu} z \right)^2
\right],
\label{e13}
\end{equation}
where the $z$ is a $N$ component complex field still subject to
the constraint $z^{\dagger} z = 1$.
The starting point of the large $N$ expansion is to take into account of the
latter constraint with help of a Lagrange multiplier field $\lambda$
with a resulting effective action:
\begin{equation}
{\cal S} = \int d^{d} x \left[
\partial_{\mu} z^{\dagger} \partial_{\mu} z-i \lambda \left(z^{\dagger} z -
\Lambda^{d-2}\frac{N}{2t}\right)
+ \frac{xt}{2N} \Lambda^{2-d}\left( z^{\dagger}
\stackrel{\leftrightarrow}{\partial}_{\mu} z \right)^2
\right]\
\label{e14}
\end{equation}
with $t=f_1$ and $x={4f_1/f_2}$.
In the absence of the current-current interaction, quartic in  $z$,
one could have integrated exactly over the $z$ field
to obtain an effective, non local, action for the $\lambda$ field from
which results the well known $1/N$ expansion of the $O(2N)$ model.
The difficulty with the massive $CP^{N-1}$ model comes precisely from
the $U(1)$ current-current interaction. The standard prescription
for dealing with such a term is to introduce a Lagrange
field ${\cal A}_{\mu}$ as in the pure $CP^{N-1}$ model$^{\cite{dadda,witten}}$.
In this case, the resulting action is given by:
\begin{equation}
{\cal S} = \displaystyle{\int d^{d} x \left[ z^{\dagger}\left(-{\cal
D}_{\mu}^{2}
- \ i \lambda \right) z+ \Lambda^{d-2}\frac{N M^2}{2} {\cal A_{\mu}}^{2}+ i
\Lambda^{d-2}\frac{N}{2t} \lambda \right]}
\label{e15}
\end{equation}
where $M^2={(1-x)/xt}$ and ${\cal D}_{\mu}= \partial_{\mu}+i{\cal A_{\mu}}$
is the covariant derivative introduced in section(2).
Now the integration over the $z$ field can be exactly  done
and the resulting partition function writes in terms of
$\lambda$ and ${\cal A}_{\mu}$ fields as:
\begin{equation}
Z = \int {\cal D} \lambda {\cal D} {\cal A_{\mu}}\ e^{\displaystyle{- N {\cal
S}_0}}
\label{e8}
\end{equation}
with:
\begin{equation}
{\cal S}_0 = {\hbox{Tr}}\ln (-{\cal D}_{\mu}^{2}
- \ i \lambda) + \Lambda^{d-2}\frac{i}{2t} \int d^{d} x\  \lambda
+ \Lambda^{d-2}\frac{M^2}{2} \int d^{d} x \ {\cal A_{\mu}}^{2}\ .
\label{e16}
\end{equation}

The set of eqs.(\ref{e8},\ref{e16}) have been obtained first
by Campostrini and Rossi$^{\cite{campos6}}$
 who have studied the resulting $1/N$ expansion in dimension $d=2$.
They have computed the mass gap and the $\beta$ functions for the
coupling constants
$t$ and $x$ at leading order in $1/N$. As a result, they argued that the mass
term $M^2$  does not renormalize, a  result which contradicts our two loop
expressions of eq.(\ref{betatx}). The origin of this contradiction between the
weak coupling and large $N$ analysis lies in the fact that the decoupling  of
the current-current interaction is not an innocent procedure.
In particular, the use of the field theoretical approach on the resulting field
theory (\ref{e16}) to obtain the correct evolution of the two coupling
constants makes sense only in a regime of energy much greater than any other
mass scale of  the theory, i.e. in
the vinicity of the $CP^{N-1}$ fixed point. This fact was missed
by Campostrini and Rossi. As a consequence, their results for the mass gap or
the
$\beta$ functions are not correct in the large mass $M^2$ limit. In this case
indeed, one has to return to the form (\ref{e14}) of the action and considerate
the current-current term
$\displaystyle \Lambda^{2-d}\frac{xt}{2N} \left( z^{\dagger}
\stackrel{\leftrightarrow}{\partial}_{\mu} z \right)^2$ as an irrelevant
operator in the neighborhood of the $O(2N)$ fixed point.

In the following, we shall study action (\ref{e13}) in the large $N$ expansion
in $d=2$ and $d>2$.  Some of our results in $d=2$ have already been
obtained  by Campostrini and Rossi$^{\cite{campos6}}$.
To be self contained, we shall present them together
with new ones.

\subsection{The ${N \rightarrow \infty}$ limit}

As readily seen on eq.(\ref{e8}), the theory is exactly soluble in the limit $N
\rightarrow \infty$ since the functional integral is dominated by the saddle
point:

\begin{equation}
\left\{
\begin{array}{lll}
\displaystyle \frac{\delta S_0}{\delta \lambda} &=& 0
\\
\\
\displaystyle \frac{\delta S_0}{\delta \cal A_{\mu}} &=&0\ .
\end{array}
\right.
\label{saddle}
\end{equation}

Searching for solutions with ${\cal A_{\mu}}$ constant one
finds ${\cal A_{\mu}} = 0$ so that
the saddle point is the same as for $O(2N)$ and $CP^{N-1}$.
Eq.(\ref{saddle}) then leads to:
\begin{equation}
\int \frac{d^{d} k} {(2\pi)^{d}} \frac{1}{k^2 + m^{2}} =
\frac{\Lambda^{d-2}}{2t}
\label{e17}
\end{equation}
with $\lambda = i m^{2}$.

In dimension $d\ge 2$, the integral (\ref{e17}) is UV divergent,
and we shall use a cut-off regularization procedure.
One has to consider separately the  $d=2$ and $d>2$ cases.

$\bullet$ In $d=2$, eq.(\ref{e17}) gives the mass gap equation:
\begin{equation}
{\pi\over t}=\ln {\Lambda\over m}
\label{e18}
\end{equation}
from which follows the $\beta$ function for $t$ to leading order $(N=\infty)$:
\begin{equation}
\beta_t ={t^2\over \pi}
\label{e19}
\end{equation}
which expresses the property of asymptotic freedom in
two dimensions$^{\cite{polyakov}}$.

$\bullet$ In $d>2$, eq.(\ref{e17}) writes:
\begin{equation}
\displaystyle{{\Gamma(\displaystyle{1-{d\over 2}})\over
(4\pi)^{\displaystyle{{d/
2}}}} \ m^{d-2}={\Lambda^{d-2}\over 2}\left({1\over t}- {1\over t_c}\right)}
\label{e20}
\end {equation}
with $\displaystyle{t_c={d-2\over 2 K_d}}$ and
$\displaystyle{K_d={2\over (4\pi)^{d/2}\Gamma({d/ 2})}}$\ .
\vskip 0.5cm
\noindent
One can then deduce the $\beta$ function for $t$:
\begin{equation}
\beta_t =-(d-2)\ t\left(1-{t\over t_c}\right).
\label{e21}
\end{equation}
The  $\beta$ function for $x$ follows trivially from the fact that
$M^2$ does not renormalize at this level since
${\cal A_{\mu}}$ does not participate at the saddle point. One thus have:
\begin{equation}
\displaystyle{\beta_{M^2} =(d-2) M^2} .
\label{e22}
\end {equation}

The set of eqs.(\ref{e19}-\ref{e22}) gives a complete description
of the RG properties of the model at $N = \infty$ and is completely consistent
with those obtained in the weak coupling analysis of the previous section:

$\bullet$ In dimension $d=2$, the expressions of the mass gap and the
$\beta$ function for $t$ agree with those obtained
from the weak coupling expanded at leading order in $N$.
Moreover, the non renormalization of the mass term $M^2$ at
this order is consistent with the fact that
$M^2$ identifies with the weak coupling one loop flow
invariant at $N = \infty$ since
$\displaystyle{M^2={1-x\over x t}} = K_{\infty}^{-1}$
(see eq.(\ref{flowinvariant})).

$\bullet$ In dimension $d > 2$, there exist two non trivial fixed points with
$t=t_c$ and $M^2=0$ or $M^2=\infty$. The fixed
point with $M^2=\infty$ governs the phase transition for all models
with $M ^2\ne 0$
since $M^2$ is a relevant variable. The critical exponent $\nu^{-1}=d-2$
is trivially  the same as for the $O(2N)$ and $CP^{N-1}$ models while
the cross-over exponent $\phi = d-2$ is just given by dimensional analysis.
Both results agree with those of eqs.(\ref{crossexp}-\ref{scalcpn}) at leading
order
in $\epsilon$ and $N$.

As seen, there is a complete agreement between weak coupling
and large $N$ approach to leading order in $1/N$ and $\epsilon$.
However, the $N=\infty$ limit is special since it does not involve
the $U(1)$ current-current interaction which is
crucial in the $CP^{N-1}$ model. This interaction
will express itself as soon as  $1/N$ corrections to
the saddle point will be taken into account and
will reveal the non trivial structure of the RG properties of the model.

\subsection{The 1/N corrections: general discussion}

Let us now investigate the effects of fluctuations around the saddle
point. The Feynman rules for the $1/N$ expansion of action (\ref{e15}) follow
from those
of the pure $CP^{N-1}$ model (see for instance Ref.{\cite{dadda}).
They are summerized in Fig.3.
%\vskip 4cm

The inverse propagator of the $\lambda$ field is given as in the $CP^{N-1}$
model
by:
\begin{equation}
\Gamma_{\lambda}(k) =
\int \frac{d^{d} q} {(2\pi)^{d}} \frac{1}{(q^2+m^2)((q+k)^2+m^2)}\ .
\label{e-pro8}
\end{equation}
The only difference with the $CP^{N-1}$ model comes from the inverse gauge
field
propagator $\cal{A_{\mu}}$ which acquires a longitudinal part:
\begin{equation}
\Gamma_{\mu \nu}(k)=\left(\delta_{\mu\nu}-{k_{\mu}k_{\nu}\over
k^2}\right)\Gamma_{\bot}(k)+\Lambda^{d-2}M^2 {k_{\mu}k_{\nu}\over k^2}
\label{e-pro10}
\end{equation}
with:
\begin{equation}
\Gamma_{\bot}(k)=\Lambda^{d-2}M^2-{4\over
d-1}{\Gamma(\displaystyle{2-{d/2}})\over (4\pi)^{\displaystyle{d/2}}}\
m^{d-2}+{k^2+4m^2\over d-1}\Gamma_{\lambda}(k)\ .
\label{e-pro11}
\end{equation}

With help of the rules of Fig.3, the $1/N $ corrections can be derived
in a standard way for all Green's functions
of the theory with the restriction that no internal spinon loop is
allowed since they are already taken into account in the definition of the
$\lambda$ propagator.
Before turning to the practical computations of the $1/N$ corrections to the
mass
gap $m$ and to the $\beta$ functions for $t$ and $M^2$ or $x$, it is worth at
this stage to precise from eq.(\ref{e-pro11}) what is the expected domain of
validity of
the decoupling procedure. To see this, let us consider the short distance
behavior of the inverse gauge field propagator involved in the computation of
the
correlation functions. In the large $k$ limit, one has:
\begin{equation}
\left\{
\begin{array}{lll}
\displaystyle{\Gamma_{\bot}(k)\build{\sim}_{k\to \infty}^{} M^2-{1\over
\pi}+{1\over 2\pi}{\ln} {k^2\over m^2}}\ \hspace{5.2 cm}{\hbox{in}}\  d=2
\\
\\
\displaystyle{\Gamma_{\bot}(k)\build{\sim}_{k\to \infty}^{}
\Lambda^{d-2}M^2-{4\over d-2}{\Gamma (2-d/2)\over (4\pi)^{d/2}} m^{d-2}+A_d\
k^{d-2}}\  \hspace{0.9 cm}{\hbox{in}}\ d>2
\end{array}
\right.
\label{asymptotic}
\end{equation}
with $ \displaystyle{A_{d} = \frac{ K_{d}}{2(d-1) S_{d}}} $ and $
\displaystyle{S_{d} = \frac{ \Gamma (d-1) \sin \left[\displaystyle{{{\pi
(d-2)}/{2}}}\right]}{2 \pi \left( \Gamma ({\displaystyle{d/2}}) \right)^{2}}}\
. $

It appears from expression (\ref{asymptotic}) that the scaling behavior of the
coupling
constants $(x,t)$ can be extracted from the UV behavior of the field theory
(\ref{e16}) when:
\begin{equation}
\left\{
\begin{array}{lll}
\displaystyle{M^2\ll {\frac{1}{2\pi}}{\ln} {\Lambda^2\over m^2} }\ \ \
 \hspace{1cm}{\hbox{in}}\ \ \ \ d=2  \\
\\
\displaystyle{M^2 \ll A_d}\ \ \ \ \ \ \hspace{1.6cm}{\hbox{in}}\ \ \ \ d>2\ .
\end{array}
\right.
\label{e25}
\end{equation}
This defines the energy regime (I) where the decoupling procedure is
meaningful:
\begin{equation}
\left\{
\begin{array}{lll}
\displaystyle{p \gg m \exp{(\pi M^2)}}\ \ \  \hspace{0.7cm}{\hbox{in}}\ \ \ \
d=2
\\
\\
\displaystyle{p \gg \Lambda \left({\frac{M^2}{A_d}}\right)^{1/d-2} }\ \ \
\hspace{0.6cm}{\hbox{in}}\ \ \ \ d>2\ .
\end{array}
\right.
\label{e251}
\end{equation}
Expanding  the second equation to leading order in $\epsilon$, one
sees that the bounds defined in eq.(\ref{e251}) identify
with those obtained in the weak coupling analysis -
eqs.(\ref{ximass},\ref{Ximasseps}) - that define the regime governed by the
$CP^{N-1}$ fixed point. Using the saddle point equations (see
eqs.(\ref{e18},\ref{e20})),
one can translate eq.(\ref{e25}) in terms of $t$ and $M^2$ as:
\begin{equation}
\left\{
\begin{array}{lll}
\displaystyle{t M^2\ll 1 }\ \ \  \hspace{2.5cm}{\hbox{in}}\ \ \ \ d=2  \\
\\
\displaystyle{t M^2 \ll {\frac{d-2}{4 \left(d-1\right) S_d}} }\ \ \ \ \ \ \
{\hbox{in}}\ \ \ \ \ d>2\ .
\end{array}
\right.
\label{e252}
\end{equation}

There exists a second regime (II) where it is possible to extract the
renormalization
of $t$  and $x$. This is the regime governed by the $O(2N)$ fixed point defined
by:
\begin{equation}
\left\{
\begin{array}{lll}
\displaystyle{p \ll m \exp{(\pi M^2)}}\ \ \   \hspace{0.8cm}{\hbox{in}}\ \ \ \
d=2
\\
\\
\displaystyle{p \ll \Lambda {\left(\frac{M^2}{A_d}\right)^{1/d-2} } }\ \ \ \ \
\ \ \ \ {\hbox{in}}\ \ \
\ d>2\
\end{array}
\right.
\label{e26}
\end{equation}
which, in terms of  $t$ and $M^2$, writes:

\begin{equation}
\left\{
\begin{array}{lll}
\displaystyle{t M^2\gg 1 }\ \ \  \hspace{2.5cm}{\hbox{in}}\ \ \ \ d=2  \\
\\
\displaystyle{t M^2 \gg {\frac{d-2}{4 \left(d-1\right) S_d}} }\ \ \ \ \ \ \
{\hbox{in}}\ \ \ \ \ d>2\ .
\end{array}
\right.
\label{e2528}
\end{equation}
In this regime the $\cal{A_{\mu}}$ field does not propagate anymore and the
decoupling (\ref{e15}) is of no use to determine the renormalization of $t$
and $x$.
One has to return to the original form of the action (\ref{e14}).
Near the $O(2N)$ fixed point, the
current-current $U(1)$ interaction is irrelevant and can be considered as
a composite operator in the $O(2N)$ theory so that its effect on the physics
can be computed perturbatively in a double expansion in $x$ and
$1/N$$^{\cite{chubukov1,chubukov2}}$.

In the following, we shall consider perturbation theory in the
regimes (I) and (II) separately.  We shall compute the $\beta$ functions for
the coupling constants $x$ and $t$ and the mass
gap $m$ at leading order in $1/N$ in $d=2$ and $d>2$.
Comparison with the results obtained from the weak coupling analysis of the
preceeding section of this work will also be made.

\subsection{The ${\bf CP^{N-1}}$ regime}

This is the regime (I)  defined by eq.(\ref{e25}). As already quoted above,
perturbation theory, as given by the Feynman rules of Fig.3, is well defined.
We shall assume in the following that the theory
is renormalizable so that only renormalization of spinon and gauge field
masses and fields are needed. We shall
thus concentrate on the two point functions of the spinon and gauge field at
order ${1/N}$:
\begin{equation}
\left\{
\begin{array}{lll}
G^{(2)}_{1/N}(p) = \langle {z^{\dagger}}(0)z(p) \rangle\\
\\
G^{(2)}_{\mu,\nu {1/N}}(p) = \langle {\cal{A_{\mu}} }(0) {\cal{A_{\nu}}}
(p)\rangle\ .
\end{array}
\right.
\end{equation}

The renormalization procedure is standard. As usual, we shall consider
the leading order of the Taylor expansion near $p=0$ of $G^{(2)}_{1/N}(p)$
and $G_{\mu,\nu {1/N}}^{(2)}(p)$ which suffer from divergent corrections which
will be eliminated
by appropriate redefinitions of the masses and fields.

\subsubsection{The two point spinon function}

The diagrams that contribute to the $1/N$ correction of $G^{(2)}_{1/N}(p)$ are
given in Fig.4$^{\cite{campos6}}$.
The self energy $\Sigma (p)$ is defined by:
\begin{equation}
G^{(2)}_{1/N}(p)={1\over p^2+m^2}-{1\over N}{1\over p^2+m^2}\Sigma(p){1\over
p^2+m^2}
\label{e27}
\end{equation}
or in terms of  the 1-P.I. function:
\begin{equation}
\Gamma^{(2)}_{1/N}(p)=p^2+m^2+{1\over N}\Sigma(p)\ .
\label{e27b}
\end{equation}
After lengthly but straightforward calculations we have obtained
for  $\Sigma(p)$:
\begin{equation}
\Sigma(p)=A p^2+B m^2+O(p^4)
\label{e28}
\end{equation}
with:
\begin{equation}
\left\{
\begin{array}{lll}
A=\displaystyle{{4-d\over d}\int {d^dk\over
(2\pi)^d}{\Gamma^{-1}_{\lambda}(k)\over
(k^2+m^2)^2}}&-&\displaystyle{{4(d-1)\over d}\int {d^dk\over
(2\pi)^d}{\Gamma^{-1}_{\bot}(k)\over k^2+m^2}+{\Lambda^{2-d}\over M^2}\int
{d^dk\over (2\pi)^d}{1\over k^2+m^2}}\\
\\
B=\displaystyle{3 \int {d^dk\over (2\pi)^d}{\Gamma^{-1}_{\lambda}(k)\over
(k^2+m^2)^2}}&+& \displaystyle{\left({\Lambda^{2-d}\over
M^2}+(d-2)(d+1){\Gamma^{-1}_{\lambda}(0)\over m^2}\right)\int {d^dk\over
(2\pi)^d}{1\over k^2+m^2}}\\
\\
&+&\displaystyle{(d-1)\left(4-(d-1)\Lambda^{d-2}M^2{\Gamma^{-1}_{\lambda}(0)\over m^2}\right)\int {d^dk\over (2\pi)^d}{\Gamma^{-1}_{\perp}(k)\over k^2+m^2}}\ .
\end{array}
\right.
\label{e29}
\end{equation}
Defining:
\begin{equation}
\Gamma_R^{(2)}(p)=Z \Gamma^{(2)}_{1/N}(p)
\label{e30}
\end{equation}
with the following prescriptions:
\begin{equation}
\left\{
\begin{array}{lll}
\displaystyle{\Gamma_R^{(2)}(p^2=0)=m_R^2}
\\
\\
\displaystyle{{\partial \Gamma_R^{(2)}(p)\over \partial p^2}\bigg |_{p=0}=1}
\end{array}
\right.
\label{e31}
\end{equation}
we obtain:
\begin{equation}
\left\{
\begin{array}{lll}
\displaystyle{Z=1-{A\over N}}
\\
\\
\displaystyle{m_R^2=m^2\left(1+{B-A\over N}\right)}.
\end{array}
\right.
\label{e32}
\end{equation}

We shall now consider the cases $d=2$ and $d>2$ separately.

\noindent $\bullet$ {\sl Two dimensional case.}

In $d=2$, we find for $Z$ and $m_R^2$:
\begin{equation}
\left\{
\begin{array}{lll}
\displaystyle{Z=1-{1\over N} \left[ {1\over 2} \ln \ln {\Lambda^2 \over m^2} -
\ln(2 \pi M^2
-2 + \ln{\Lambda^2 \over m^2}  ) + {1\over 4 \pi M^2} \ln {\Lambda^2 \over m^2}
\right]}
\\
\\
\displaystyle{m_R^2=m^2\left[1+{1\over N}\left( \ln \ln {\Lambda^2 \over
m^2} + (3-2 \pi M^2)
\ln(2 \pi M^2 -2 + \ln{\Lambda^2 \over m^2}  ) \right) \right]}\ .
\end{array}
\right.
\label{e331}
\end{equation}
These expressions agree with those of Ref.\cite{campos6}.
Using the saddle point
equation (\ref{e18}), one can express the mass gap $m_R^2$ as a
function of $t$ and $M^2$:
\begin{equation}
\displaystyle{m_R^2=m^2\left[1+{1\over N}\left( \ln {2 \pi \over t} + (3-2 \pi
M^2)
\ln(2 \pi M^2 -2 + {2 \pi \over t}  ) \right) \right]}\ .
\label{e333}
\end{equation}
Using the $N=\infty$ result (\ref{e19}) and  the fact
that $\beta_{M^2}= O(1/N)$ since $M^2$ does not renormalize at the saddle point
one obtains the $\beta$ function for $t$:
\begin{equation}
\displaystyle \beta_t={t^2\over \pi}\left[1+{t\over 2\pi N}\left(1+{3-2\pi
M^2\over 1+t\displaystyle(M^2-{1/ \pi})}\right)\right].
\label{e34}
\end{equation}
This result has been first
obtained by Campostrini and Rossi$^{\cite{campos6}}$.

\noindent $\bullet$ {\sl Above two dimensions.}

Let us begin by giving the explicit expressions of the field and mass
renormalizations in $d=3$:

\begin{equation}
\left\{
\begin{array}{lll}
\displaystyle{Z=1-{1\over N} \left[ {4\over 3\pi^2} \ln {\Lambda\over m} +
{1\over 2 t M^2}-{64\over 3 \pi^2} \ln \left({\Lambda \over 16}
+M^2\Lambda-{\displaystyle {m}\over \displaystyle{2\pi}}\right) \right]}
\\
\\
\displaystyle{m_R^2=m^2\left[1+{32\over3\pi^2  N}\ln {\Lambda \over m} +
{16\over \pi N} {\Lambda\over m}+ {256\over \pi^2 N}\left({1\over 3}-{\pi
\Lambda M^2\over m}\right) \ln \left({\Lambda \over 16}
+M^2\Lambda-{\displaystyle {m}\over \displaystyle{2\pi}}\right) \right]}\
\end{array}
\right.
\label{e38}
\end{equation}
which leads to the  $\beta$ function for $t$:
\begin{eqnarray}
\beta_t=-\left[t-{t^2\over \pi^2}\left(1+{4\over N}-{4\over N}{M^2\over
{1/16}+M^2+{1/t}-{1/ \pi^2}}\right)\right] \nonumber \\
\left[1+{16\over 3N\pi^2}+{8\over 3N\pi^2}{1\over {1/16}+M^2+{1/t}-{1/
\pi^2}}\right]\ .
\label{e39}
\end{eqnarray}
In $d>2$, we can obtain the $\beta$ function for $t$ from eqs.(\ref{e20}) and
(\ref{e29}):

\begin{equation}
\begin{array}{clc}
\displaystyle\beta_t = &- \displaystyle \left[
t -  {t^{2}\over t_c} \left( 1 + \frac{d^{2}-d-2}{
N} - \frac{(d-1)^{2}}{N} \frac{M^2}{A_{d}+ M^2 +
\displaystyle{1/t} -{1/t_c} } \right) \right] \\
\\
& \displaystyle \left[ d-2 + \frac{4(d-1)(d-2) S_{d}}{Nd}
+ \frac{2(d^{2}-1)(d-2) K_{d}}{Nd} \frac{1}{A_{d}+
M^2 + \displaystyle{1/t} - {1/t_c}}\right]\ .
\end{array}
\label{e35}
\end{equation}

\subsubsection{Gauge field renormalization}

We shall now investigate the role of fluctuations on the gauge field
$\cal{A_{\mu}}$ at order $1/N$ in two and three dimensions.
This will provide the renormalization of the gauge field mass $M^2$ from which
will follow
the renormalization of the  complete set of coupling constants of action
(\ref{e16}). From the
renormalization point of view, the situation is formally equivalent to what
happens in Quantum Electrodynamics. In particular, Ward identities stated in
the treatment of gauge field with a non-vanishing mass for the photon ensure
the absence of a counter-term for the longitudinal part of the gauge field
propagator in such a way that the renormalization constant for the gauge field
$\cal{A_{\mu}}$, $Z_{\cal{A_{\mu}}}$ and for its mass, $Z_{M^2}$ verify
$Z_{\cal{A_{\mu}}}Z_{M^2}=1$. It follows from this that the inverse gauge field
propagator at order ${1/N}$ can be written:
\begin{equation}
\Gamma_{\mu \nu {1/N}}(p)=\left(\delta_{\mu\nu}-{p_{\mu}p_{\nu}\over
p^2}\right)\Gamma_{\bot  {1/N}}(p)+M^2\ {p_{\mu}p_{\nu}\over p^2}
\label{eq1}
\end{equation}
where $\Gamma_{\bot{1/N}}(p)$ is the transversal part of the inverse gauge
field
propagator at order $1/N$.

We define the renormalized inverse gauge field propagator by:
\begin{equation}
\Gamma_{\mu \nu R}(p)=\left(\delta_{\mu\nu}-{p_{\mu}p_{\nu}\over
p^2}\right)\Gamma_{\bot R}(p)+M_R^2\ {p_{\mu}p_{\nu}\over
p^2}
\label{eqr}
\end{equation}
with:
\begin{equation}
\left\{
\begin{array}{lll}
\displaystyle{\Gamma_{\bot R}(p)=Z^{-1}_{\cal{A_{\mu}}}\Gamma_{\bot{1/N}}(p)}
\\
\\
\displaystyle{M_R^2=Z^{-1}_{\cal{A_{\mu}}}M^2}
\end{array}
\right.
\label{eq2}
\end{equation}
where $Z_{\cal{A_{\mu}}}$ is determined by the prescriptions:
\begin{equation}
\displaystyle{\Gamma_{\bot R}(p=0)=M_R^2}
\label{presc}
\end{equation}
and:
\begin{equation}
\left\{
\begin{array}{lll}
\displaystyle{{\partial \Gamma_{\bot R}\over \partial p^2}\bigg |_{p=0}={1\over
12\pi m_R^2}\ \hspace{2.5 cm}{\hbox{in}}\ d=2}\
\\
\\
\displaystyle{{\partial \Gamma_{\bot R}\over \partial p^2}\bigg |_{p=0}={1\over
24\pi m_R}\ \hspace{2.5 cm}{\hbox{in}}\ d=3}\ .
\end{array}
\right.
\label{prescription}
\end{equation}

The diagrams contributing to the renormalization of the gauge field
$\cal{A_{\mu}}$ at this order can be found in
Refs.\cite{campos6,campos3,campos4} and
they are reproduced in Fig.5.
The  renormalization of $\Gamma_{\bot}(p)$ is provided by a Taylor expansion in
power of the external moment $p$ of the diagrams. We then need the
expressions of the transversal part of the bare inverse gauge field propagator
at small
$p$:
\begin{equation}
\left\{
\begin{array}{lll}
\displaystyle{\Gamma_{\bot}(p)\build{\sim}_{p\to 0}^{}{{{p^2}}\over
{12\pi{m^2}}}  + M^2 + {{{\hbox{O}}(p^4)}}} \ \hspace{2.5 cm}{\hbox{in}}
\ d=2
\\
\\
\displaystyle{\Gamma_{\bot}(p)\build{\sim}_{p\to 0}^{}{{{p^2}}\over {24\pi
m}}  + M^2 + {{{\hbox{O}}(p^4)}} \ \hspace{2.7 cm}}{\hbox{in}} \ {d=3}
\end{array}
\right.
\label{eqi}
\end{equation}
which, taking into account of the spinon mass renormalization at order $1/N$
given by eq.(\ref{e32}), leads to:
\begin{equation}
\left\{
\begin{array}{lll}
\displaystyle{\Gamma_{\bot}(p)\build{\sim}_{p\to 0}^{}{{{p^2}}\over
{12\pi{m_R^2}}}  + {p^2\over 12 \pi m_R^2} {B-A\over N}+ M^2
+{{{\hbox{O}}(p^4)}}} \ \hspace{2.5 cm}{\hbox{in}} \ {d=2}
\\
\\
\displaystyle{\Gamma_{\bot}(p)\build{\sim}_{p\to 0}^{}{{{p^2}}\over {24\pi
m_R}} + {p^2\over 48 \pi m_R}{B-A\over N} + M^2  +{{{\hbox{O}}(p^4)}} \
\hspace{2.5 cm}}{\hbox{in}} \ \ {d=3}\ .
\end{array}
\right.
\label{eq2e}
\end{equation}
The calculation of the counter-terms for the inverse gauge field propagator at
order
$1/N$ is very cumbersome and details are given in the Appendix.

We have obtained the following expressions:

\begin{equation}
\left\{
\begin{array}{lll}
\displaystyle{\Gamma_{\bot{1/N}}(p)=\Gamma_{\bot}(p)}-{p^2\over 12 \pi
m_R^2}{B-A\over N} \ \hspace{2.5 cm}{\hbox{in}} \ {d=2}
\\
\\
\displaystyle{\Gamma_{\bot{1/N}}(p)=\Gamma_{\bot}(p)}-{p^2\over 48 \pi
m_R}{B-A\over N} \ \hspace{2.5 cm}{\hbox{in}} \ {d=3}
\end{array}
\right.
\label{eq4}
\end{equation}
which, together with  eq.(\ref{eq2e}), show that the counter-terms associated
with the transversal part of the inverse gauge field propagator are just
cancelled by
the renormalization of the spinon mass in such a way that the gauge field
renormalization is trivial:
\begin{equation}
Z_{\cal{A_{\mu}}}=1\ .
\label{eq413}
\end{equation}
We thus find that the gauge field mass $M^2$ {\sl does not renormalize} at
order
$1/N$ in $d=2$ and $d=3$. This result probably holds
for any dimension between $2$ and $4$. However, it is worth stressing
that this result is true only in the small $tM^2$ limit where
the above calculation makes sense.
The non renormalization of $M^2$ at order $1/N$ and the relation between $M^2$
and $x$ - $M^2={(1-x)/xt}$ - induce a non trivial renormalization for the
variable $x$ given by:
\begin{equation}
\beta_x=-{M^2\over (1+tM^2)^2}\ \beta_t\ .
\label{betax}
\end{equation}

\subsubsection{Comparison between large $N$ and  weak coupling  analysis}

At first sight, the $\beta$ functions for $t$ and $x$ seem
to interpolate between the $CP^{N-1}$ and $O(2N)$ models: they behave smoothly
as  functions of $tM^2$. The $\beta$ function (\ref{e34}) for $t$ allows to
recover that of the $CP^{N-1}$ and $O(2N)$ models as $M^2$ goes to zero or
infinity respectively. However, as discussed above, the whole calculation and,
in particular, that involving gauge field loops is valid in a narrow
range of energy defined by $tM^2\ll 1$ which also  implies  $1-x\ll1$.
We shall now see that it is only in that case
that the results of the ${1/N}$ approach agree with those of the weak coupling
analysis. In contrast,
we shall show that the naive extrapolation of the results for large
$tM^2$ disagrees with the weak coupling ones.

Let us first  show that there is, in the $CP^{N-1}$ regime, a good agreement
with the weak coupling results of the preceeding section. We consider now
separately the cases
$d=2$ and $d > 2$.

Let us consider the $\beta$ functions (\ref{betatx}) obtained in the weak
coupling
analysis in two dimensions. These equations can be expanded in powers of $1/N$
by doing the substitution $T\to {t/N}$. The resulting expressions can then be
expanded in
powers of $1-x$. Keeping only terms of first order in this parameter one
obtains
the following expressions:
\begin{equation}
\left\{
\begin{array}{lll}
{\displaystyle \beta_t}&\hspace{-0.3cm}=&\hspace{-0.3cm}
\displaystyle \frac{t^2}{\pi}\left(1 -{1\over N} +
{x\over N}\right) + \displaystyle \frac{2t^3}{\pi^2 N}\\
&&\\
{\displaystyle \beta_x}& \hspace{-0.3cm}=&\hspace{-0.3cm}
- \displaystyle \frac{t}{\pi}\ (1-x)-\frac{2t^2}{\pi^2 N}(1-x)\ .
\end{array}
\right .
\label{betatxnn}
\end{equation}
Now one can formally expand the $\beta$ functions (\ref{e34}) and
(\ref{betax}) obtained in the $1/N$ approach in powers of $t$ up to order
$t^3$. Then $M^2$ is replaced by its expression as a function of $x$ and $t$.
Keeping again just terms of order $1-x$, we recover the set of equations
(\ref{betatxnn}).

This matching between the large $N$ and weak coupling expansions also works for
the mass gap. The expression (\ref{e333}) of the mass gap obtained in the large
$N$ analysis
can be expanded in powers of $t$ and re-exponentiated in powers of $1/N$. This
leads to:
\begin{equation}
\xi =C \Lambda^{-1} \  {\displaystyle {t^{{\displaystyle{\frac{2}{N}}}}}}\
\exp{\left({\displaystyle{{\pi\over {t}}}}\right)} \exp \
\left({\displaystyle{-\frac{\left(1-x\right)\pi}{Nt}}{\ln}\frac{t}{2\pi}}\right)
\label{xixi}
\end{equation}
which coincides at leading order in $t$ with the expression (\ref{xiCPN}) of
$\xi$ obtained in the weak coupling analysis in the $CP^{N-1}$ regime.

In dimension greater than two, one can compare the exponents $\nu_{CP^{N-1}}$
and
$\phi_{CP^{N-1}}$ that govern the leading scaling behavior of $x$ and $t$ at
the $CP^{N-1}$ fixed point. The exponent $\nu_{CP^{N-1}}$  obtained in the
$1/N$ expansion from eq.({\ref{e35})$^{\cite{hikami1}}$:
\begin{equation}
\displaystyle \nu_{CP^{N-1}}={1\over d-2}\left(1-{4d(d-1)S_d\over N}\right)
\label{e36}
\end{equation}
coincides with that obtained in the weak coupling analysis (eq.(\ref{nucpn}))
at  order
$1/N$ and $\epsilon^2$. The cross-over exponent $\phi_{CP^{N-1}}$
can be obtained in the large $N$ limit from eq.(\ref{betax}) -
$\phi_{CP^{N-1}}= d-2$ - in agreement with the weak coupling result.
We can therefore conclude that, in the $CP^{N-1}$ regime, the large $N$ and
weak coupling expansions coincide.

Such an agreement does not exist in the $O(2N)$ regime. In two dimensions, let
us consider the  $\beta$ functions obtained in the weak coupling analysis and
re-expanded in
powers of ${1/N}$:
\begin{equation}
\left\{
\begin{array}{lll}
{\displaystyle \beta_t}&\hspace{-0.3cm}=&\hspace{-0.3cm}
\displaystyle \frac{t^2}{\pi}\left(1 -{1\over N} +
{x\over N}\right) + \displaystyle \frac{t^3}{2\pi^2 N}\\
&&\\
{\displaystyle \beta_x}& \hspace{-0.3cm}=&\hspace{-0.3cm}
- \displaystyle \frac{xt}{\pi}\left(1+{t\over \pi N}\right).
\end{array}
\right .
\label{betatxn5}
\end{equation}
These expressions have to be compared to the $\beta$ functions obtained in the
large $N$ analysis expanded in powers of ${1/tM^2}$ since the condition
$tM^2\gg 1$ defines the $O(2N)$ regime:
\begin{equation}
\left\{
\begin{array}{lll}
{\displaystyle \beta_t}=
\displaystyle \frac{t^2}{\pi}\left(1-{1\over N} +
{x\over N}\right) + \displaystyle \frac{t^3}{2\pi^2 N}(1+x)\\
&&\\
{\displaystyle \beta_x}=-\displaystyle
\frac{xt}{\pi}\left(1-\frac{1}{N}\right)-\displaystyle \frac{xt^2}{2\pi^2N}\ .
\end{array}
\right .
\label{betatxn6}
\end{equation}
As readily seen, the two set of equations (\ref{betatxn5}) and (\ref{betatxn6})
differ for the $\beta$ function for $x$ even at {\it first} order in $t$.
This disagreement persists, of course, for the mass gap by  a term which is
linear in $x/N$.

In dimension greater than two, let us compare the critical exponents
obtained from the $\beta$ functions of eqs.(\ref{e35}, \ref{betax}) with
those obtained in the weak coupling expansion.

The critical exponent $\nu_{ O(2N)}$ as obtained  from
eq.(\ref{e35})$^{\cite{ma2}}$:
\begin{equation}
\displaystyle \nu_{O(2N)}={1\over d-2}\left(1-{4(d-1)S_d\over Nd}\right)\
\label{e37}
\end{equation}
coincides with that obtained in the weak coupling expansion (eq.(\ref{nuo2n}))
at order $1/N$ and $\epsilon^2$.
However, since  the gauge field mass does not renormalize at order ${1/N}$,
the cross-over exponent is given by dimensional analysis:
$\phi_{O(2N)}=d-2+O({1/N^2})$, and thus disagrees with the weak coupling
expression of
eq.(\ref{crossexp}).
There is, in the $O(2N)$ regime, a disagreement between the $1/N$ expansion
performed with help of the decoupling procedure and the weak coupling
expansion. The reason is that the decoupling procedure does not take
properly into account the dangerously irrelevant character of the
current-current operator. The agreement we have obtained for the critical
exponent $\nu_{O(2N)}$ stems from the fact that the exponent $\nu$ is
determined
by the asymptotic scaling in the relevant direction $t$ at $x = 0$.

Naturally, this  disagreement  was expected since as previously discussed,
the decoupling procedure is valid provided $tM^2 \ll 1$.
We shall now see that, provided one   performs a suitable  $1/N$ expansion in
the vinicity of the $O(2N)$ fixed point - i.e. when $tM^2 \gg 1$-, the weak
coupling and the large $N$ analysis are consistent.

\subsection{The ${\bf O(2N)}$ regime}

In this regime,  the gauge
field ${\cal A}_{\mu}$ no longer propagates and perturbation theory defined by
the Feynman rules of Fig.3 is meaningless. In particular, the decoupling of
the current-current term
$\left(z^{\dagger}\stackrel{\leftrightarrow}{\partial}_{\mu} z \right)^2$ is of
no use for exploring this low energy regime and one has to come
back to the original form of action (\ref{e14}). Simple power counting around
the $O(2N)$ theory indicates that
$\left(z^{\dagger}\stackrel{\leftrightarrow}{\partial}_{\mu} z \right)^2$ has
scaling dimension $2-d$ and is therefore an irrelevant (resp. marginal
irrelevant)
operator in dimension $d > 2$ (resp. in $d=2$). Its effects on the low energy
physics can however be computed in a double expansion in $x$ and $1/N$
around the
$O(2N)$ theory  as soon as $x\ll 1$ and $N\gg
1$$^{\cite{chubukov1,chubukov2}}$.
In the following, we shall
compute the corrections to the RG functions at first order in $x$ and
$1/N$.
The Feynman rules for the double expansion follow from those of the
$1/N$ expansion of the $O(2N)$ sigma model. Propagators of the spinon $z$ and
$\lambda$ fields as well as the $\lambda\bar{z} z$ vertex
are unchanged. The only modification is the presence
of a new  four point
vertex $\Gamma_{\mu}$ which has been first introduced
by  Chubukov et al.$^{\cite{chubukov1,chubukov2}}$ in their study of the
quantum version of the model in $d=2+1$ (see Fig.6):
\begin{equation}
\Gamma_{\mu} \left(k_1,k_2,k_3,k_4\right) = \Lambda^{2-d} \frac{xt}{2N}
\left(k_1+k_3\right)_{\mu} \left(k_2+k_4\right)_{\mu}
z^{\dagger}_{\alpha} \left(k_1\right) z^{\dagger}_{\beta} \left(k_2\right)
z_{\alpha} \left(k_3\right) z_{\beta} \left(k_4\right).
\label{vert4part}
\end{equation}
This operator being irrelevant we expect that higher order
derivatives irrelevant operators will be generated by renormalization.
However as we are only interested in the leading correction in $x$ we can omit
these possible additional terms.

As in the preceeding subsection, we shall consider the renormalization of the
spinon two point function and the  four point vertex function
$\Gamma_{\mu}^{(4)}$ at first order in $x$ and $1/N$ from which the RG
functions
will be computed and finally compared with those previously obtained
in the weak coupling analysis.

\subsubsection{The two point spinon function}

The diagrams contributing to the
two point function $G^{(2)}_{1/N}(p)$ at first order in $x$ and $1/N$
are depicted in Fig.7. The  coefficients $A$ and $B$  entering in the
expression of the self energy $\Sigma(p)$  of eq.(\ref{e28})  are now given by:
\begin{equation}
\left\{
\begin{array}{lll}
\displaystyle A = \frac{4-d}{d} \int \frac{d^dk}{(2\pi)^d}
\frac{\Gamma^{-1}_{\lambda}(k)}{\left(k^2+m^2\right)^2} &-& xt \Lambda^{2-d}
\displaystyle \int \frac{d^dk}{(2\pi)^d} \frac{1}{k^2+m^2}\\
\\
\displaystyle B = 3 \int \frac{d^dk}{(2\pi)^d}
\frac{\Gamma^{-1}_{\lambda}\left(k\right)}{\left(k^2+m^2\right)^2} &-& \left(
\displaystyle \left(3-d\right) \frac{\Gamma^{-1}_{\lambda}\left(0\right)}{m^2}
+ \Lambda^{2-d} xt\right) \displaystyle \int \frac{d^dk}{(2\pi)^d}
\frac{1}{k^2+m^2}\\
\\
&+&
\displaystyle x t \Lambda^{2-d}  \Gamma^{-1}_{\lambda}\left(0\right)
\left(\int \frac{d^dk}{(2\pi)^d} \frac{1}{k^2+m^2}\right)^2\ .
\end{array}
\right.
\label{doubletwo}
\end{equation}
In terms of this quantities, the field and mass renormalizations
are given by eq.(\ref{e32}).

$\bullet$ In two dimensions, we find:
\begin{equation}
\left\{
\begin{array}{lll}
\displaystyle{Z=1-{1\over N} \left[{1\over 2} \ln \ln {\Lambda^2 \over m^2}
- \frac{xt}{4\pi} \ln {\Lambda^2 \over m^2} \right]}
\\
\\
\displaystyle{m_R^2=m^2\left[1+{1\over N}\ln \ln {\Lambda^2 \over
m^2}
- {1\over N}\ln{\Lambda^2 \over m^2} +{1\over N} \frac{xt}{4\pi}
\left(\ln{\Lambda^2 \over m^2}\right)^2 \right]} \ .
\end{array}
\right.
\label{ed331}
\end{equation}
With use of  the saddle point equation (\ref{e18}), the mass gap $m_R^2$
writes in terms of $x$ and $t$ as:
\begin{equation}
\displaystyle{m_R^2=m^2\left[1+{1\over N}\left(\ln {2\pi \over t}- {2\pi \over
t}+{x\pi \over t}\right)\right]}\
\label{ed3311}
\end{equation}
from which follows the $\beta$ function for $t$:
\begin{equation}
\beta_t = \frac{t^2}{\pi}\left(1- \frac{1}{N}+\frac{x}{N}\right)
+ \frac{t^3}{2\pi^2 N}\ .
\label{ed333}
\end{equation}

$\bullet$ In dimension greater than two, the divergent part of $A$ and $B$
are easily computed and, omitting the details, we find for $\beta_t$:
\begin{equation}
\beta_t = t^2 \left[ \left(d-2\right) \left(\frac{1}{t_c} - \frac{1}{t}
\right) \left(1 + \frac{4\left(d-1\right)}{Nd} S_d \right)
+ \frac{\left(d-2\right)\left(d-3\right)}{Nt_c} +
\frac{x\left(d-2\right)}{Nt_c} \right]\ .
\label{prett}
\end{equation}

\subsubsection{The four point vertex function}

The expression of $\beta_{x}$ can be obtained by considering
the renormalization of the vertex function at first order in
$x$ and $1/N$. At this order, the four point vertex function is
given using the diagrams of Fig.8 by:
\begin{equation}
\Gamma_{\mu {1/N}}^{(4)}\left(k_1,k_2,k_3,k_4\right) =
-\frac{\Lambda^{2-d}xt}{2N} \left(k_1+k_2\right)_{\mu}
\left(k_3+k_4\right)_{\mu}
\left[ 1 - \frac{6}{N} {d-2\over d} \int \frac{d^dk}{(2\pi)^d}
\frac{\Gamma^{-1}_{\lambda}(k)}{\left(k^2+m^2\right)^2} \right] \ .
\label{vertebcor}
\end{equation}
The renormalized vertex function is defined through:
\begin{equation}
\Gamma_{\mu R}^{^{(4)}}\left(k_1,k_2,k_3,k_4\right) =
Z^2 \ \Gamma_{\mu {1/N}}^{(4)}\left(k_1,k_2,k_3,k_4\right)
\label{vertedef}
\end{equation}
where $Z$ is the spinon field renormalization. At this order  we just need the
field renormalization $Z$ of the $O(2N)$ model.  Defining $Z_{xt}$ by:
\begin{equation}
(xt)_R=Z_{xt}\  xt\
\end{equation}
we deduce from eq.(\ref{vertebcor}):
\begin{equation}
Z_{xt} =  1 -\frac{4\left(d-1\right)}{Nd}
\int \frac{d^dk}{(2\pi)^d}
\frac{\Gamma^{-1}_{\lambda}(k)}{\left(k^2+m^2\right)^2}\
\label{coup}
\end{equation}
and explicitly:
\begin{equation}
\left\{
\begin{array}{lll}
\displaystyle{Z_{xt} =   1 -\frac{1}{N}
\ln \ln \frac{\Lambda^2}{m^2}}  \hspace{4 cm} {\hbox{in}} \ \ d=2
\\
\\
\displaystyle{Z_{xt}=  1 -\frac{8\left(d-1\right)}{Nd}
S_d \ln \frac{\Lambda}{m}}  \hspace{3 cm}    {\hbox{in}}  \ \ d>2\ .
\end{array}
\right.
\label{coupd}
\end{equation}
Using these results, together with the expressions of the $\beta$ functions
for $t$ given by eqs.(\ref{ed333},\ref{prett}),  we deduce the
$\beta$ function for the variable $x$:
\begin{equation}
\left\{
\begin{array}{lll}
\displaystyle{\beta_x = -\frac{xt}{\pi}\left(1+\frac{t}{2\pi N}\right)}
\hspace{6 cm}{\hbox{in}}\ \ d=2
\\
\\
\displaystyle{\beta_x = - \left[d-2+\frac{8\left(d-1\right)}{Nd} S_d +
{{\beta_{t}^{O(2N)}} \over t} \right] \ x}
\hspace{2.5 cm}{\hbox{in}} \ \ d>2\ ,
\end{array}
\right.
\label{betaxxx}
\end{equation}
where in the last equation, $\beta_t^{O(2N)}$ is the $\beta$ function for $t$
for the
$O(2N)$
model.

\subsubsection{Comparison between weak coupling and large $N$ analysis}

In dimension two, one sees that the  $\beta$ functions of
eqs.(\ref{ed333},\ref{betaxxx})  agree, at first order in $x$ and ${1/N}$, with
the  weak coupling expressions of eq.(\ref{betatxn5}).
There is still a disagreement for the two loop term of the $\beta$ function for
$x$ but
 this  corresponds to subdominant
corrections to the scaling behavior at the $O(2N)$ fixed point. Anyway,
this term does not contribute to the mass gap at this order. Indeed,  one can
check
that the  expression of the correlation length obtained from
eq.(\ref{ed3311}):
\begin{equation}
\displaystyle{\xi=C' \Lambda^{-1} t^{\displaystyle{1\over 2(N-1)}}\
\exp\left({\displaystyle{\pi N\over t(N-1)}}\right) \exp\left({\displaystyle{-
x \pi\over 2Nt}}\right)}
\label{ed3312}
\end{equation}
identifies with
the correlation length (eq.(\ref{xio2N})) obtained in the weak coupling
expansion.

In dimension greater than two,  the cross-over exponent $\phi_{O(2N)}$ is
obtained from
eq.(\ref{betaxxx}):
\begin{equation}
\phi_{O(2N)} = d-2 + \frac{8\left(d-1\right)}{Nd} S_d\ .
\label{crossexpn}
\end{equation}
For example, in $d=3$, we find $\phi_{O(2N)} = 1 + 32/3\pi^2 N$,
in agreement with the result of Chubukov et al$^{\cite{chubukov1,chubukov2}}$.
As  seen, $\phi_{O(2N)}$ is not given by dimensional analysis but displays a
non trivial
correction at order $1/N$ contrarily to what was found in the preceeding
subsection.
This is the result  of the important fact that, in the $O(2N)$ regime, the
gauge field mass $M^2$ {\sl does}
renormalize contrarily to what asserted by Campostrini and
Rossi$^{\cite{campos6}}$.  Finally, it is easy to show that
eq.(\ref{crossexpn})
identifies, at order $1/N$ and $\epsilon^2$ with the weak coupling result
(eq.(\ref{crossexp})):
\begin{equation}
\phi_{O(2N)} = \epsilon + \frac{\epsilon}{N} + \frac{\epsilon^2}{2N}\ .
\label{crossexpneps}
\end{equation}

Altogether these results show that the double expansion in $1/N$ and $x$
agrees  with the weak coupling analysis provided one is sufficiently closed to
the $O(2N)$ fixed point.

\section{CONCLUSION}

In this work we have  studied  the
${SU(N)\otimes U(1)/SU(N-1)\otimes U(1)}$ NL$\sigma$ model which is relevant
for the low energy physics of frustrated spin systems. We have shown that this
model interpolates between two different fixed points which are
$O(2N)$ and $CP^{N-1}$  symmetric. Our main goal, in this paper, was to inquire
the reliability of perturbation theory in such a multi-fixed point situation
about which little is known at present.
To this end, we have performed a rather complete analysis of the RG properties
of
the model using the weak coupling and $1/N$ expansions at next to leading
order.

The main characteristic of frustrated systems is the presence of an additional
length
scale $\Xi$ that determines
the scaling regimes governed by the different fixed points of the theory.
While the
 presence of this scale does not alter the weak coupling expansion, it has
strong
 consequenses for the $1/N$ expansion. Indeed, the standard approach of the
$1/N$ expansion
 fails to explore all the phase diagram: the decoupling procedure used to
tackle with the
 $U(1)$ current-current interaction is meaningless as one removes from the
basin of
 attraction of the $CP^{N-1}$ fixed point. We have shown that another strategy
has to be
 used to explore the physics for large values of the gauge field mass $M^2$:
around the
 $O(2N)$ fixed point a double expansion in $1/N$ and $x$ {\sl is necessary} to
 obtain the correct renormalization of the parameters of the massive $CP^{N-1}$
model. Doing this, the results obtained from weak coupling and $1/N$ expansions
agree in the very neighborhood of the two fixed points of the theory, giving us
confidence
 on perturbation theory in the vicinity of these fixed points. However this
procedure also
 addresses the question of universality for frustrated systems and for models
where
 several antagonistic fixed points coexist. Indeed, the weak coupling expansion
of
 NL$\sigma$ models, as we know from the work of Friedan, is renormalizable.
This implies
 that a priori only two coupling constants $x$ and $T$ are sufficient to
determine the low
 energy behavior when $\xi^{-1}\ll p\ll \Lambda$, i.e. in the whole range of
variation of
 $x$, $0\le x \le 1$. In this sense we expect the behavior of frustrated
systems to be
 universal. This scheme appears to be more questionable in view of the large
$N$ expansion since one may suspect that in the  double expansion in $1/N$ and
 $x$,  higher orders corrections in $x$ involve more and more irrelevant
operators and
 coupling constants. If it is the case the existence of a universal field
theory describing the
 whole phase diagram of frustrated systems should be ruled out and only a local
idea of
 universality, bounded to the vicinity of the fixed points, would be
meaningful.

A pratical consequence of the previous discussion is that it could be very
difficult to observe the $O(2N)$ scaling behavior when  dealing with numerical
simulations with systems of size $L\ll \Xi$. Indeed, for sizes of order $L \sim
\Xi$  one can expect to observe rather than the $O(2N)$ scaling some effective
exponents that result from the competition of $O(2N)$ and $CP^{N-1}$
fluctuations. This might be one reason why, up to now, the $O(4)$ critical
behavior has not been observed in the Monte Carlo simulations of frustrated
Heisenberg spin systems. A better understanding of finite size scaling in the
model is in progress.

{\bf Acknowledgements}:

We are pleased to thank M. Caffarel, B. Delamotte, F. Delduc,
V. Fateev, P.K. Mitter, and S. Sachdev for useful discussions.
\newpage

{\bf Appendix : Gauge field renormalization.}

We give here the asymptotic UV behavior  of the diagrams  contributing to the
renormalization of the transversal part of the inverse gauge field propagator
$\Gamma_{\bot}(p)$ in 2 and 3 dimensions. Since there is no renormalization of
the longitudinal part of the gauge field propagator we can forget, in the
diagrams of Fig.5 involving internal loop of the gauge field, the longitudinal
part $Si_{\parallel}$ and keep only their transversal part $Si_{\bot}$.
Moreover to determine the renormalization of the propagator, we can restrict to
search for the counter-term proportional to $p^2$ since it defines completely
the renormalization of the propagator itself.

$\bullet$ In $d=2$, we have obtained:
\begin{equation}
\begin{array}{lll}
\displaystyle S_1=p^2\left({-1\over 12 \pi m_R^2}\ln \ln {\Lambda^2\over
m_R^2}\right)\\
\\
\displaystyle S_2=p^2\left({1\over 24 \pi m_R^2}\ln \ln {\Lambda^2\over
m_R^2}+{1\over 48 \pi m_R^2}
{\hbox{li}}{\Lambda^2\over m_R^2}\right)\\
\\
\displaystyle S_3=p^2\left({-1\over 24 \pi m_R^2} \ln {\Lambda^2\over
m_R^2}-{1\over 48 \pi m_R^2}
{\hbox{li}}{\Lambda^2\over m_R^2}+{1\over 24 \pi m_R^2}\ln \ln {\Lambda^2\over
m_R^2}\right)\\
\\
\displaystyle S_{4\bot}=0\\
\\
\displaystyle S_{5\bot}=p^2\left({-1\over 24 \pi m_R^2} \ln \left (2\pi
M^2-2+\ln {\Lambda^2\over m_R^2}\right)\right)\\
\\
\displaystyle S_{6\bot}=p^2\left({1\over 24 \pi m_R^2} \ln {\Lambda^2\over
m_R^2}-{M^2 \over 12 m_R^2}\ln \left (2\pi M^2-2+\ln {\Lambda^2\over
m_R^2}\right)+{1\over 12 \pi m_R^2}\ln \left (2\pi M^2-2+\ln {\Lambda^2\over
m_R^2}\right)\right)\\
\\
\displaystyle S_{7\bot}=p^2\left({1\over 24 \pi m_R^2}\ln \left (2\pi M^2-2+\ln
{\Lambda^2\over m_R^2}\right) \right)\ .
\end{array}
\end{equation}

The global contribution to $\Gamma_{\bot}(p)$ at order ${1/N}$ thus writes:
\begin{equation}
\begin{array}{lll}
\displaystyle S_1+2S_2+2S_3+S_{4\bot}+2S_{5\bot}+2S_{6\bot}+4S_{7\bot}= \\
\\
\displaystyle p^2 \left({1\over 12\pi m_R^2}\ln \ln \displaystyle
{\Lambda^2\over m_R^2}+{1\over 12 \pi m_R^2}\left(3-2\pi M^2\right)\ln \left
(2\pi M^2-2+\ln {\Lambda^2\over m_R^2}\right)
\right)
\end{array}
\end{equation}
and therefore we have:
\begin{equation}
\displaystyle{\Gamma_{\bot{1/N}}(p)=\Gamma_{\bot}(p)}-{p^2\over 12 \pi
m_R^2}{1\over N}\left (\ln \ln {\Lambda^2\over m_R^2}+(3-2\pi M^2)\ln \left
(2\pi M^2-2+\ln {\Lambda^2\over m_R^2}\right)\right) \ .
\end{equation}

$\bullet$ In $d=3$, the cut-off dependence of the diagrams writes:
\begin{equation}
\begin{array}{lll}
\displaystyle S_1=p^2\left({-1\over 18 \pi^3 m_R}\ln {\Lambda^2\over
m_R^2}\right)\\
\\
\displaystyle S_2=p^2\left({\Lambda \over {6 \pi^4 m_R^2}} + {\Lambda ^2\over
48\pi^3m_R^3} +{1\over {3 \pi^5 m_R}}\ln {\Lambda^2\over m_R^2}\right)\\
\\
\displaystyle S_3=p^2\left(-{\Lambda \over 6 {\pi^4}{m_R^2}} -{\Lambda ^2\over
48 \pi^3 m_R^3} - {1\over {3\pi^5m_R}}\ln {\Lambda^2\over m_R^2}+ {1\over
{12\pi^3 m_R}}
   \ln {\Lambda^2\over m_R^2}\right)\\
\\
\displaystyle S_{4\bot}=0\\
\\
\displaystyle S_{5\bot}=p^2\left({-2\over 3 \pi^3 m_R}\ln \left ({\Lambda\over
16} + M^2 \Lambda -{m\over 2\pi} \right)\right)\\
\\
\displaystyle S_{6\bot}=p^2\left({\Lambda\over 6 \pi^2 m_R^2} + {2 \over 3
\pi^3 m_R^2 }
(m_R-4\pi \Lambda M^2)\ln \left ({\Lambda\over 16} + M^2 \Lambda -{m\over
2\pi}\right)\right)\\
\\
\displaystyle S_{7\bot}=p^2\left({4\over 9 \pi^3 m_R}\ln \left ({\Lambda\over
16} + M^2 \Lambda -{m\over 2\pi}\right) \right)
\end{array}
\end{equation}
and the complete contribution to $\Gamma_{\bot}(p)$ at order ${1/N}$ is:
\begin{equation}
\begin{array}{lll}
\displaystyle S_1+2S_2+2S_3+S_{4\bot}+2S_{5\bot}+2S_{6\bot}+4S_{7\bot}=\\
\\
\displaystyle p^2\left({\Lambda\over 3 m_R^2 \pi^2} +{2\over 9 m_R \pi^3}\ln
\displaystyle {\Lambda^2\over m_R^2}+{1\over 9\pi^3 m_R^2}(16m_R-48\pi \Lambda
M^2)\ln \left ({\Lambda\over 16} + M^2 \Lambda -{m\over
2\pi}\right)\right)\nonumber
\end{array}
\end{equation}
so that:
\begin{equation}
\displaystyle{\Gamma_{\bot{1/N}}(p)=\Gamma_{\bot}(p)}-{p^2\over 48 \pi
m_R}{1\over N}\left({32\over 3\pi^2} \ln {\Lambda\over m_R}+{16\Lambda\over \pi
m_R}+{256 \over \pi^2}\left({1\over 3}-{\pi \Lambda M^2\over m_R}\right)\ln
\left ({\Lambda\over 16} + M^2 \Lambda -{m\over 2\pi}\right)\right) \ .
\end{equation}

\newpage

\newpage

\vskip 1cm

\centerline{\bf{Figure Captions}}

\vskip 1cm

Fig.1: The infrared RG flow in $d=2$ in the ($x,T$) plane.

Fig.2: The infrared RG flow in $d=2+\epsilon$ in the ($x,T$) plane.

Fig.3: Feynman rules for the $1/N$ expansion in the $CP^{N-1}$
regime. The solid line denotes spinon propagator while dashed
and wavy lines represent respectively the $\lambda$ and ${\cal A}_{\mu}$
propagators.

Fig.4: Diagrams contributing to the two point function at order $1/N$ in the
$CP^{N-1}$
regime.

Fig.5: Diagrams contributing to the gauge field propagator at order $1/N$ in
the $CP^{N-1}$ regime.

Fig.6: The four point vertex occuring in the double expansion
in $x$ and $1/N$ in the $O(2N)$ regime.

Fig.7: Diagrams contributing to the two point function at first order in
$1/N$ and $x$ in the $O(2N)$ regime.

Fig.8: Diagrams contributing to the four point vertex function at first order
in
$1/N$ and $x$ in the $O(2N)$ regime.

\vspace{-1cm}
\begin{figure}[h]
\vspace{8cm}
\hspace{4cm}
\psfig{figure=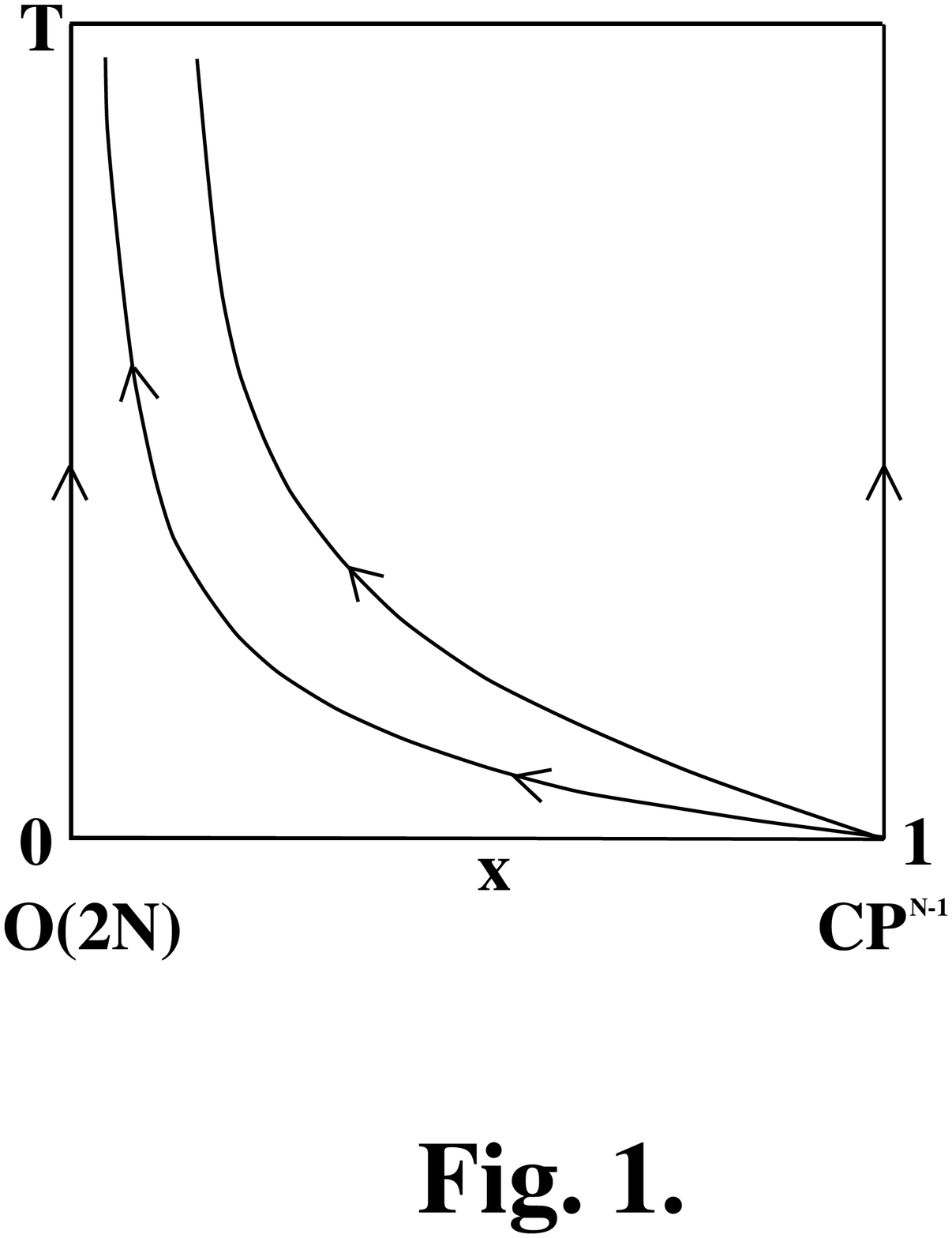,height=12cm}
\vspace{5cm}
\protect\label{F00}
\end{figure}
\vspace{1cm}


\newpage

\vspace{-1cm}
\begin{figure}[h]
\vspace{8cm}
\hspace{40cm}
\psfig{figure=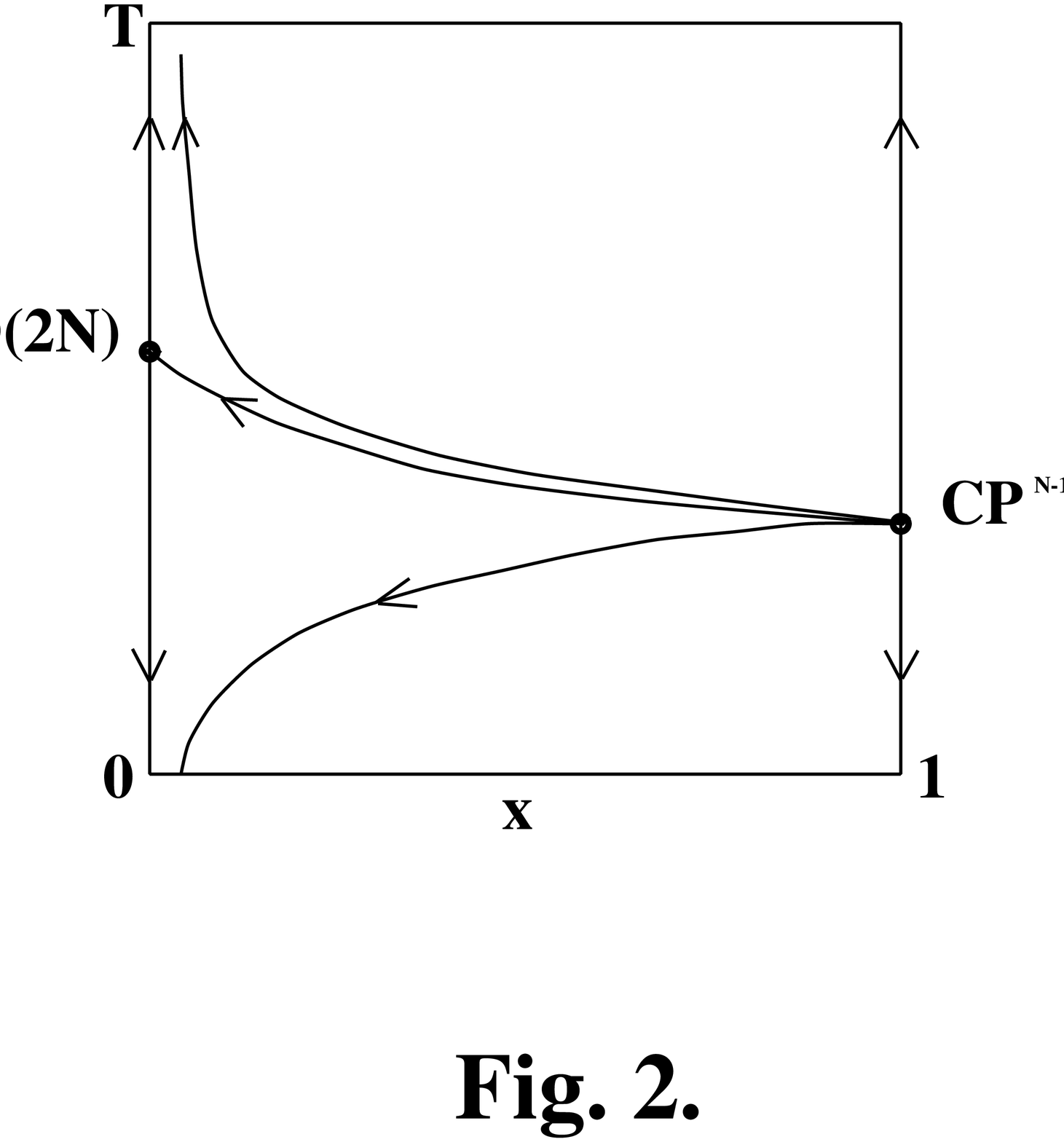,height=12cm}
\vspace{5cm}
\protect\label{F0}
\end{figure}
\vspace{1cm}


\newpage

\vspace{-1cm}
\begin{figure}[h]
\vspace{12cm}
\hspace{15cm}
\psfig{figure=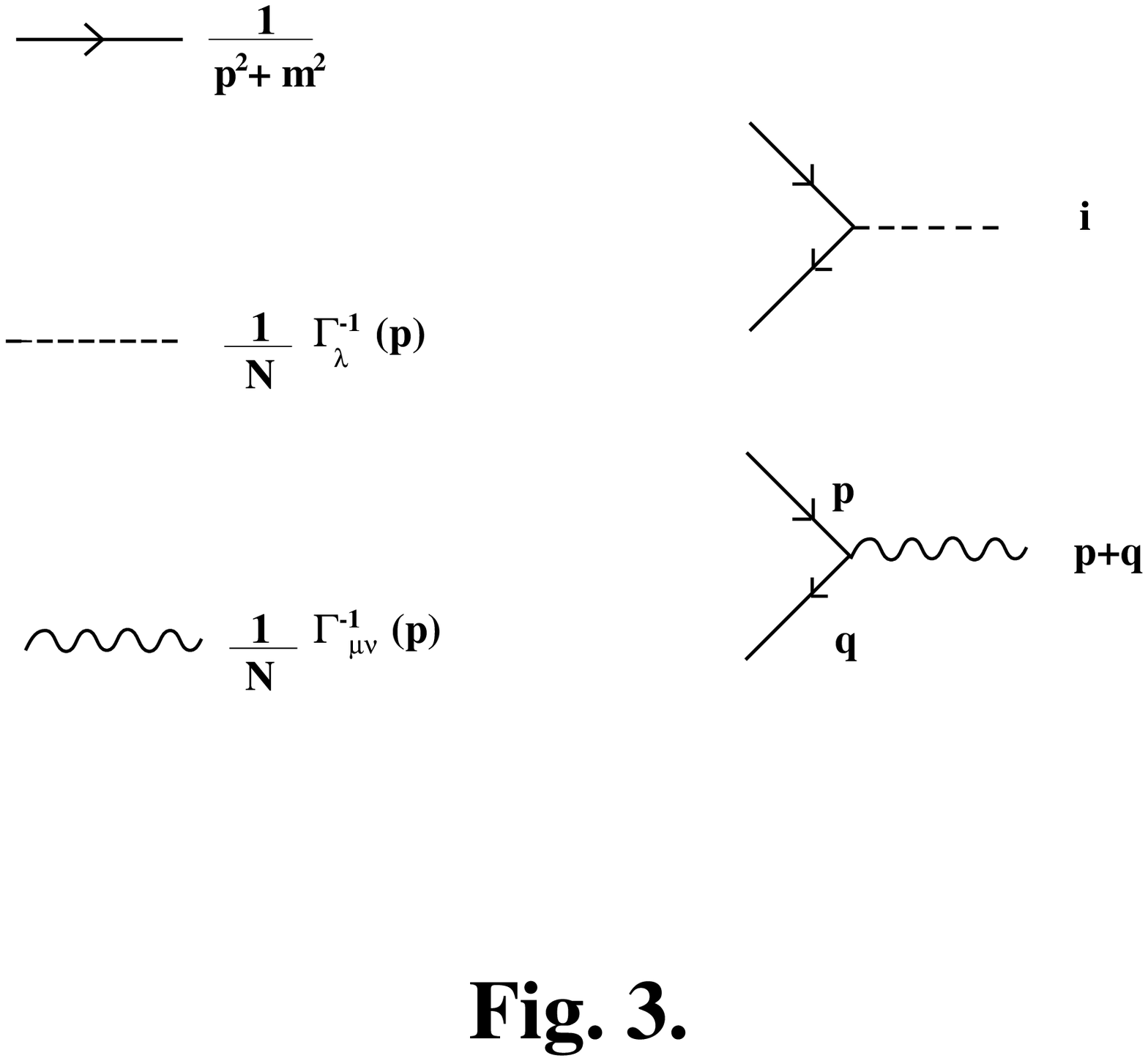,height=12cm}
\protect\label{F1}
\end{figure}


\newpage

\vspace{-3cm}
\begin{figure}[h]
\vspace{23cm}
\hspace{-3cm}
\psfig{figure=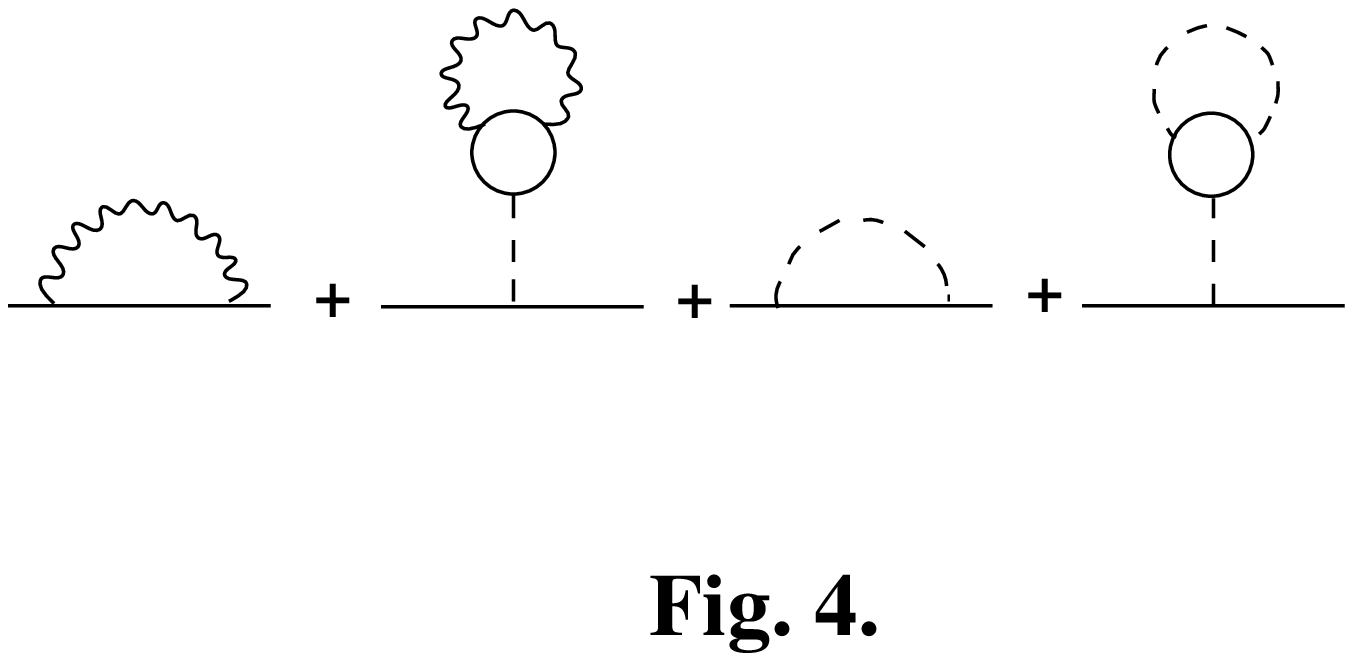,height=7cm}
\protect\label{F2}
\end{figure}


\newpage

\vspace{-3cm}
\begin{figure}[h]
\vspace{6cm}
\hspace{-2cm}
\psfig{figure=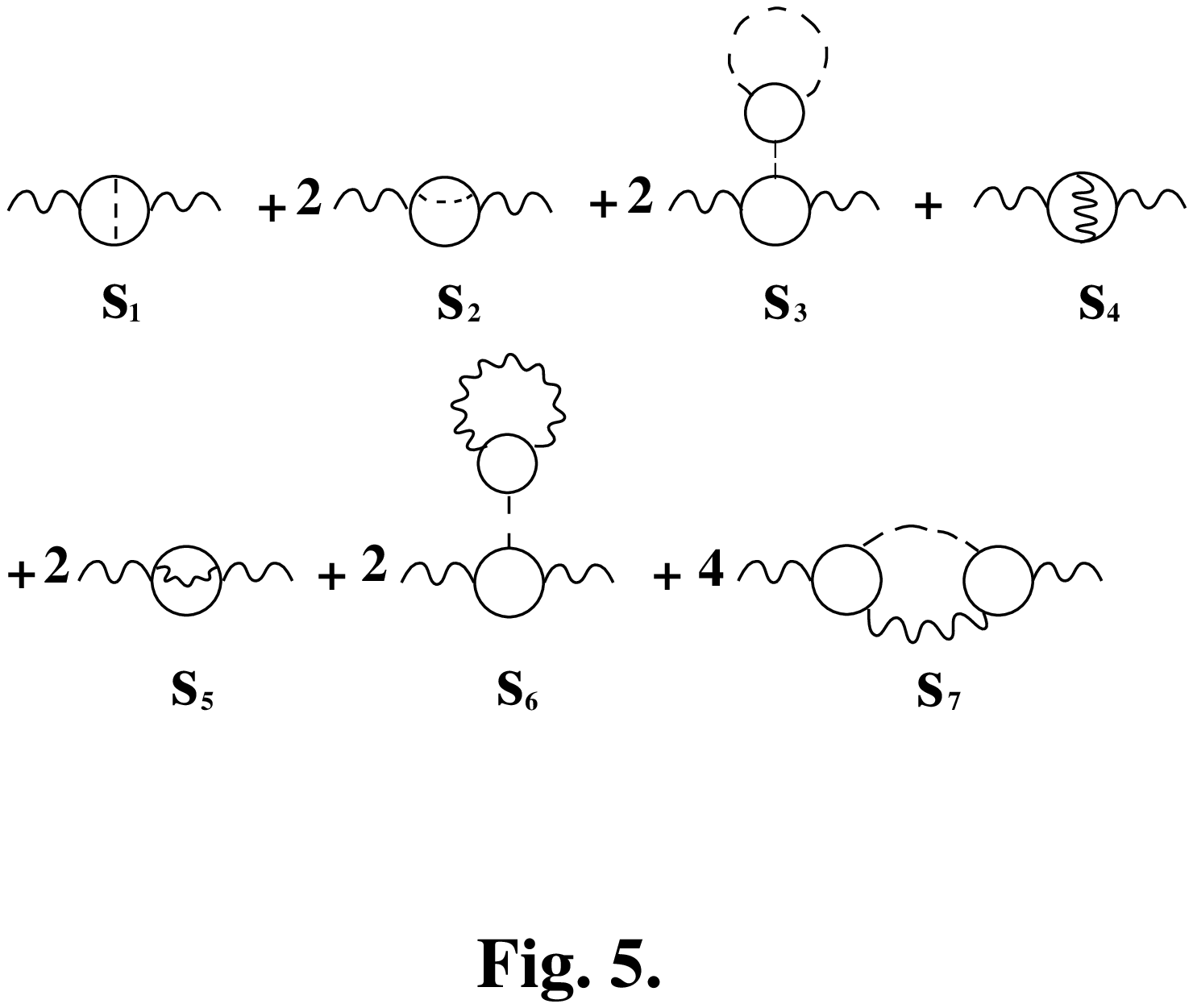,height=13cm}
\protect\label{F3}
\end{figure}


\newpage

\vspace{-1cm}
\begin{figure}[h]
\vspace{13cm}
\hspace{4cm}
\psfig{figure=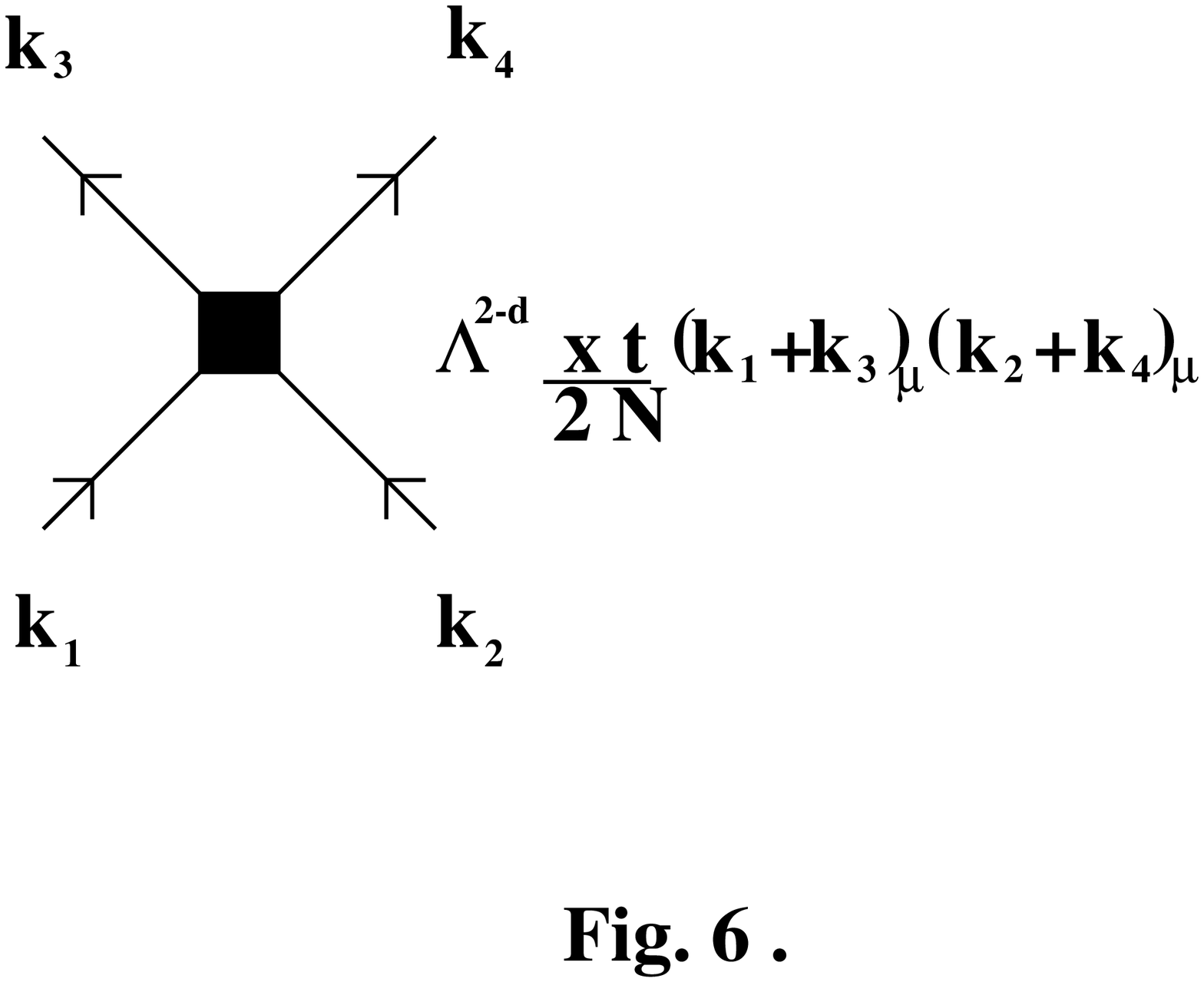,height=11cm}
\vspace{5cm}
\protect\label{F4}
\end{figure}
\vspace{1cm}


\newpage

\vspace{-1cm}
\begin{figure}[h]
\vspace{19cm}
\hspace{3cm}
\psfig{figure=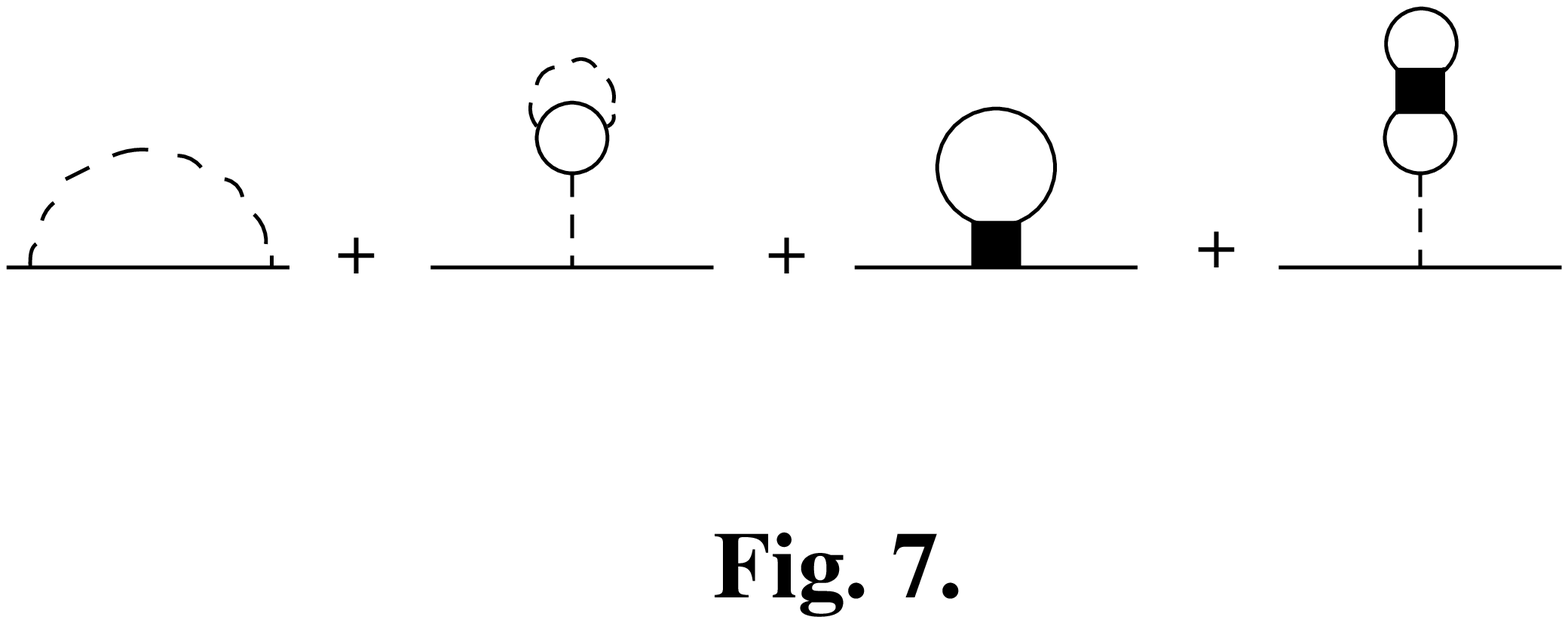,height=12cm}
\vspace{5cm}
\protect\label{F5}
\end{figure}
\vspace{1cm}


\newpage

\vspace{-1cm}
\begin{figure}[h]
\vspace{16cm}
\hspace{1cm}
\psfig{figure=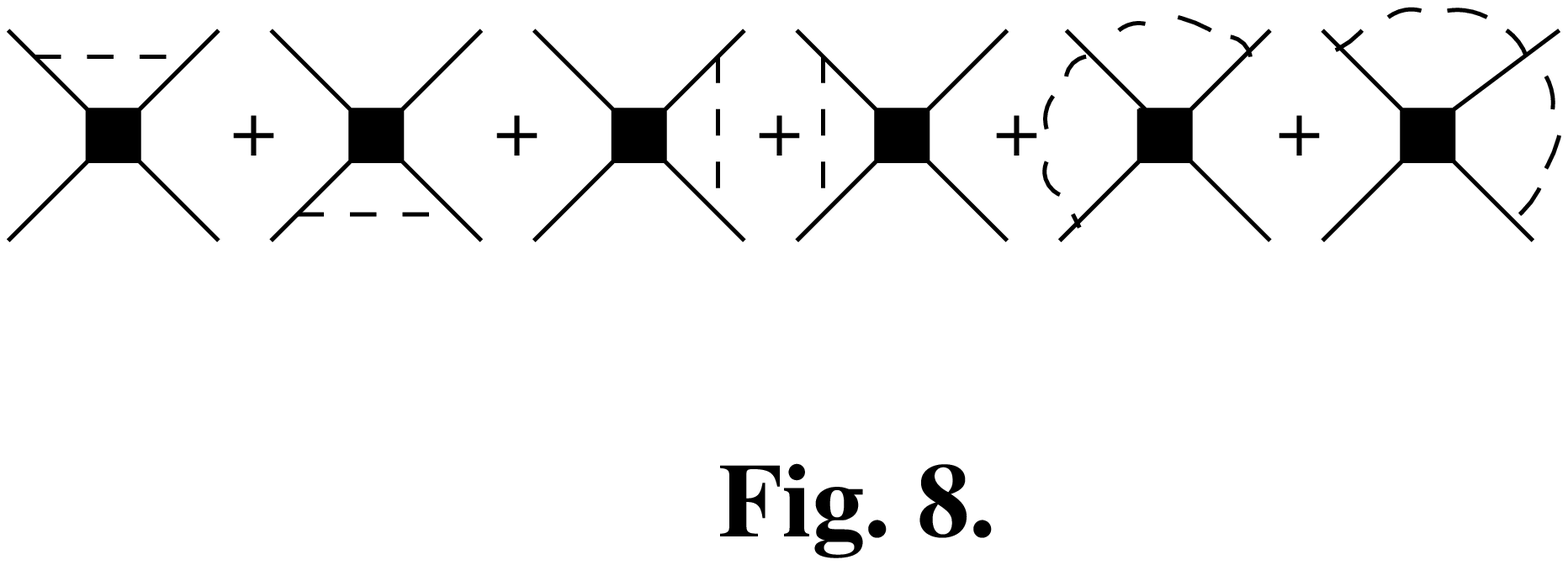,height=6cm}
\vspace{5cm}
\protect\label{F6}
\end{figure}
\vspace{1cm}


\end{document}